\documentclass [11pt, letterpaper]{article}
\pdfoutput=1 

\usepackage{fullpage}
\usepackage{amsmath}
\usepackage{amssymb}
\usepackage{graphicx}
\usepackage{mathrsfs}
\usepackage{wrapfig}
\usepackage{boxedminipage}
\usepackage{epsfig}
\usepackage{setspace}
\usepackage{subfigure}
\usepackage{float}
\usepackage{hyperref}
\usepackage{enumerate}
\usepackage{cite}
\usepackage[font=footnotesize,labelfont=rm]{caption}

\begin{document}

\begin{flushright}
{\tt arXiv:1511.08782}
\end{flushright}

{\flushleft\vskip-1.35cm\vbox{\includegraphics[width=1.25in]{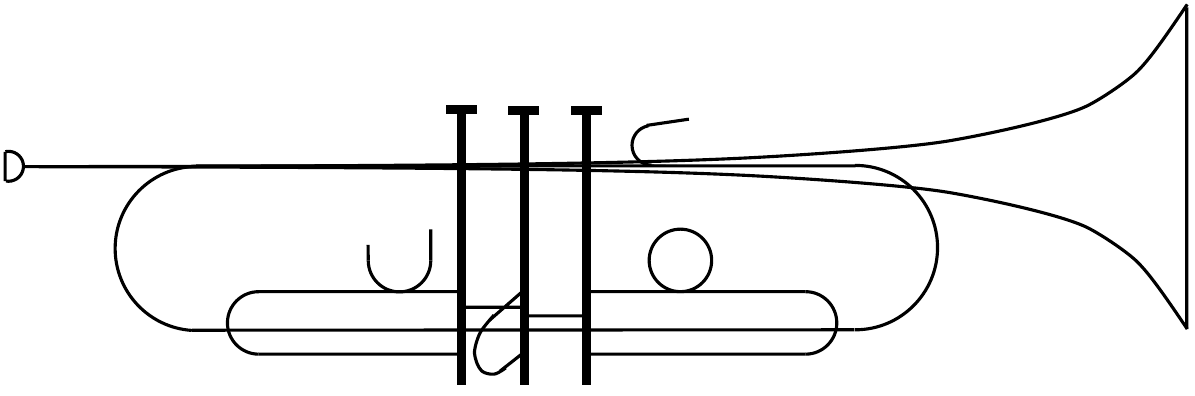}}}

\bigskip
\bigskip
\bigskip
\bigskip

\bigskip
\bigskip
\bigskip
\bigskip

\begin{center} 

{\Large\bf  Gauss--Bonnet Black Holes}

\bigskip

{\Large\bf and}

\bigskip

{\Large\bf Holographic Heat Engines Beyond Large $N$}

\end{center}

\bigskip \bigskip \bigskip \bigskip

\centerline{\bf Clifford V. Johnson}

\bigskip
\bigskip
\bigskip

  \centerline{\it Department of Physics and Astronomy }
\centerline{\it University of
Southern California}
\centerline{\it Los Angeles, CA 90089-0484, U.S.A.}

\bigskip

\centerline{\small \tt johnson1,  [at] usc.edu}

\bigskip
\bigskip


\begin{abstract} 
\noindent  Working in the extended black hole  thermodynamics where a dynamical cosmological constant defines a thermodynamic pressure $p$, we study the efficiency of heat engines that perform mechanical work {\it via} the $pdV$ terms  now present in the First Law. Here the black hole itself is the working substance, and we focus on a judiciously chosen  engine cycle. We work in Gauss--Bonnet--Einstein--Maxwell gravity with negative cosmological constant and, using a high temperature expansion, compare the results for these ``holographic'' heat engines to  that of previously studied   cases with no Gauss--Bonnet sector. From the dual holographic large~$N$ field theory perspective, this amounts to studying the effects of a class of $1/N$ corrections to the  efficiency of the cycle.
\end{abstract}
\newpage \baselineskip=18pt \setcounter{footnote}{0}

\section{Introduction}
%

By making the cosmological constant $\Lambda$  a dynamical variable in  a theory of gravity, an interesting extension of the classic black hole thermodynamics\cite{Bekenstein:1973ur,Bekenstein:1974ax,Hawking:1974sw,Hawking:1976de} can be made\footnote{For a selection of references, see refs.\cite{Caldarelli:1999xj,Wang:2006eb,Sekiwa:2006qj,LarranagaRubio:2007ut,Kastor:2009wy,Dolan:2010ha,Cvetic:2010jb,Dolan:2011jm,Dolan:2011xt}. See also the early work in refs.\cite{Henneaux:1984ji,Teitelboim:1985dp,Henneaux:1989zc}.}. 
The  cosmological constant of the spacetime in question supplies a pressure {\it via} $p=-\Lambda/8\pi$, usually missing (along with its counterpart $V$)  from the traditional framework\footnote{Here we are using geometrical units where $G,c,\hbar,k_{\rm B}$ have been set to unity. We may restore them using dimensional analysis when required later.}. While the temperature and entropy remain related to the surface gravity and area in the usual way, the mass ends up being related not to the internal energy $U$ of the system, but the {\it enthalpy}, $H$, as discussed in ref.\cite{Kastor:2009wy}:
\begin{equation}
M=H\equiv U+pV\ ,\quad T=\frac{\kappa}{2\pi}\ , \quad S=\frac{A}{4}\ ,
\end{equation}
where the volume $V$ is the conjugate of the pressure $V\equiv(\partial H/\partial p)|_S=(\partial M/\partial p)|_S$, following from the First Law which is now 
$dM=TdS+Vdp\ . 
$
The black holes may have other parameters such as gauge charges $q_i$ and angular momenta $J_i$, and these, with their conjugates the potentials  $\Phi_i$ and angular velocities $\Omega_i$, enter additively into the First Law in the usual manner. This extended black hole thermodynamics  formalism works in multiple dimensions.
(Interestingly, in the static black hole case, the thermodynamic volume $V$ is associated with the naive volume occupied by the black hole itself: The volume of the ball of radius given by the horizon radius (denoted $r_+$ here).   

It was noted in ref.\cite{Johnson:2014yja} that since  pressure and volume variables are now present alongside temperature and entropy,  a device  for extracting  useful mechanical work from heat energy --- a traditional heat engine --- may be defined. (Similarly, heat pumps or refrigerators, where instead  work  is done to transfer heat from a cold reservoir to a hot one may be defined as well.)  Such devices were dubbed ``holographic heat engines", since especially in the case of negative cosmological constant (where pressure is positive) such cycles presumably represent a journey through a family of holographically dual\cite{Maldacena:1997re,Witten:1998qj,Gubser:1998bc,Witten:1998zw,Aharony:1999t} non--gravitational large $N$ field theories defined in one dimension fewer. The physics of such engines may possibly have interesting and instructive implications for those field theory tours, which may be uncovered when the dictionary between this extended thermodynamics and holography is fully worked out, as discussed in refs.\cite{Johnson:2014yja,Johnson:2016pfa}\footnote{A recent paper\cite{Karch:2015rpa} has made some progress in working out other aspects of the dictionary, and will certainly be relevant. Thanks to Phuc Nguyen for pointing out this reference.}, and we will make some further  remarks upon this below, and in the concluding section.

The black hole defines  an equation of state, coming from the relation between the temperature, horizon radius,  other black hole parameters, and the cosmological constant. It defines (either implicitly or explicitly)  a function  $p(V,T)$, and our  engine can be defined as a closed path in the  $p{-}V$ plane, allowing for the input of an amount of heat $Q_H$, and the exhaust of an amount~$Q_C$. 
\begin{wrapfigure}{L}{0.3\textwidth}
{\centering
\includegraphics[width=1.4in]{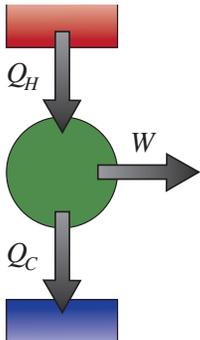} 
   \caption{\footnotesize   The  heat engine flows.}   \label{fig:heatengine}
}
\end{wrapfigure}
The total mechanical work done, by the First Law, is of course $W=Q_H-Q_C$. A central  quantity,   the efficiency of the heat engine, is defined to be $\eta= W/Q_H = 1-Q_C/Q_H$. Figure~\ref{fig:heatengine} shows the standard logic of the energy flows for one cycle of the engine.

The exact properties of the engine depends upon the equation of state (defined by the type of black hole\footnote{Refs.\cite{Belhaj:2015hha,Sadeghi:2015ksa,Caceres:2015vsa,Setare:2015yra} have since done  studies of holographic heat engines using various kinds of black holes in diverse dimensions. }) and the choice of path in the $p{-}V$ plane. Of course, the efficiency of the engine is bounded above by the Carnot efficiency given by that of a reversible heat engine made by  expanding along an isotherm, and then an adiabat, followed by contracting along an isotherm and then using an adiabat to close the path.  With the  isotherms at temperatures $T_H$ and~$T_{\rm C}$,  where $T_H>T_{\rm C}$, the Carnot efficiency is $\eta_C=1-T_{\rm C}/T_H$.

Interestingly, as also noted in ref.\cite{Johnson:2014yja}, the other simple (but in general less efficient) choice, connecting the isotherms by isochoric paths defining the Stirling cycle,  turns out to be equivalent to Carnot for static black holes. This is  because in such cases the entropy and the volume both depend on the same single variable, the horizon radius, and are therefore not independent\cite{Dolan:2010ha}. So isochors and adiabats are identical in such cases making Stirling equivalent to Carnot.

 \begin{wrapfigure}{r}{0.43\textwidth}
{\centering
\includegraphics[width=2.3in]{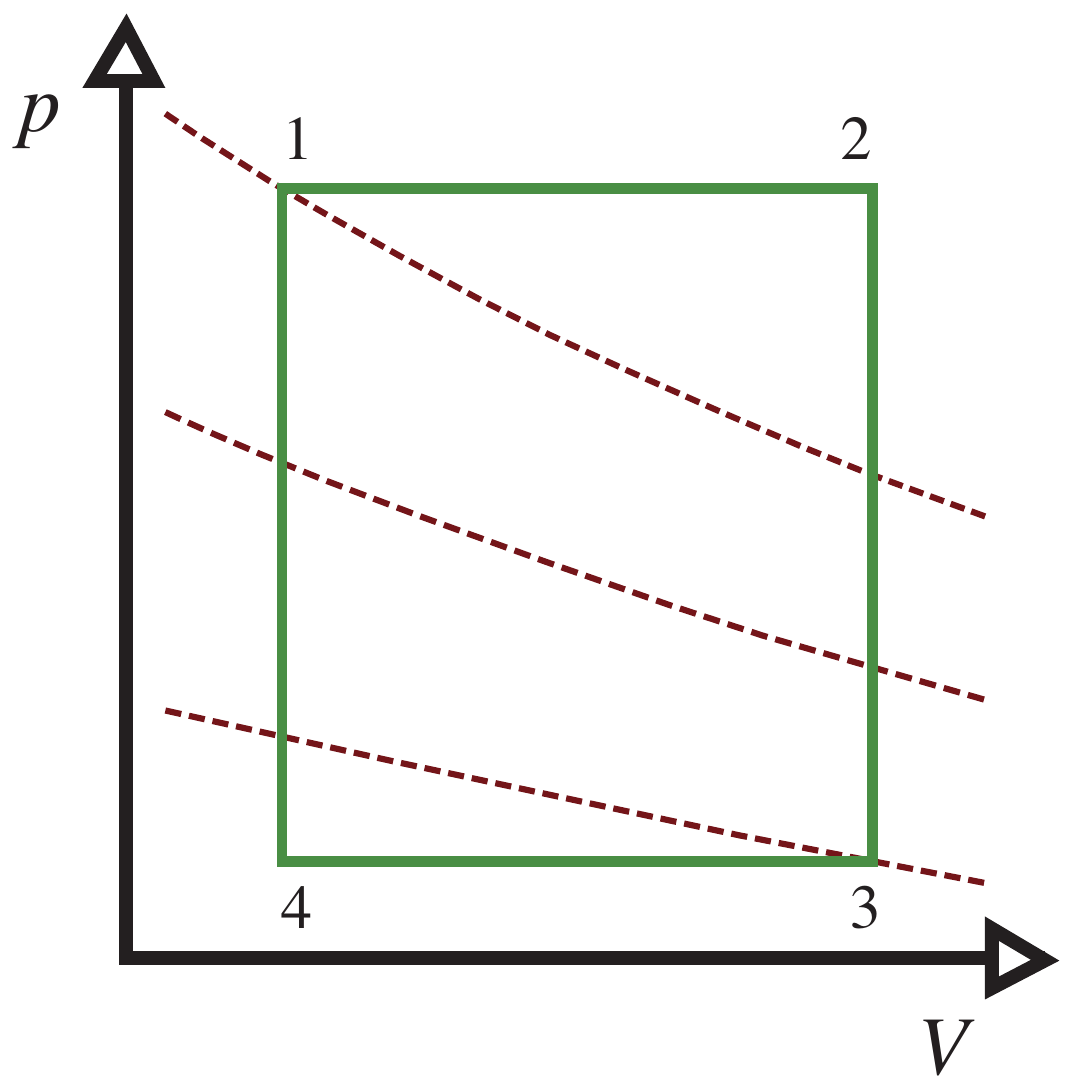} 
   \caption{\footnotesize  Our engine.}   \label{fig:cyclesb}
}
\end{wrapfigure}
That last observation is also equivalent to the fact that the specific heat at constant volume, $C_V$, vanishes for static black holes. On the other hand,~$C_p$ can be quite explicitly  computed in terms of $r_+$ and hence the entropy or the  volume. This is useful because it means that another cycle in the $p{-}V$ plane  is natural to think about: A rectangle composed of  isobars and isochors. The heat flows and hence the efficiency is in principle readily computable if one has a suitable expression for $C_p$. Looking at figure \ref{fig:cyclesb},  the work done along the isobars is:
 \begin{equation}
 W= \left(V_2-V_1\right)(p_1-p_4)\ ,
 \label{eq:nicework}
 \end{equation}
 where the subscripts refer to the quantities evaluated at the corners labeled (1,2,3,4). The heat flows take place  along the top and bottom, with the upper isobar  giving the  net inflow of heat,  which is therefore $Q_H$, and so:
 \begin{equation}
 Q_H=\int_{T_1}^{T_2} C_p(p_1,T) dT\ ,
 \label{eq:hothothot}
\end{equation}
 The efficiency is then $\eta=W/Q_H$. To get an explicit expression requires one to be able to write the specific heat in terms of $T$ in order to do the integral. This turns out to be difficult to do exactly, since the natural variable to get $C_p$ in terms of is $r_h$, and turning that into a $T$ dependence is somewhat messy. It is at this point that one can move to a more tractable regime for the problem, and do a high temperature expansion. In the next section we will  review how that works somewhat more thoroughly than was explored in the original reference, also providing new results. This is preparation for  the generalisation we will explore in this paper, the case of black holes in Gauss--Bonnet gravity\cite{Zwiebach:1985uq}, with $\Lambda<0$ \cite{Cai:2001dz}. 
 
A Gauss--Bonnet action coupled to an Einstein--Hilbert--Maxwell sector will be our focus, with the lowest non--trivial dimension being $D=5$ since in lower dimensions the Gauss--Bonnet action is purely topological. Adding a Gauss--Bonnet action to the Einstein--Hilbert--Maxwell action (with $\Lambda<0$) is interesting and important to consider since generically, whatever the parent theory of quantum gravity may be, there will be higher order corrections to the pure Einstein sector, and Gauss--Bonnet  is  a particular combination of terms that allows for tractable systematic investigation of the effects of such corrections. Additionally,  such terms represent part of the $1/N$ correction to the large~$N$ limit of the holographically dual $SU(N)$--like gauge field theory (see ref.\cite{Aharony:1999t} for a review). So while this study is interesting in its own right, exploring the properties of holographic heat engines after including corrections to the leading order  gravity theory, there is also the possibility that we learn something about  how their properties are adjusted  away from the large $N$ limit, using Gauss--Bonnet as a laboratory, in the spirit of {\it e.g.} refs.\cite{Brigante:2007nu,Brigante:2008gz}. 

We will be able to answer the question (at least in the high temperature limit) as to whether the key quantity, the efficiency of the heat engine, increases or decreases in the presence of the Gauss--Bonnet terms. It is not {\it a priori} obvious what the answer should be.  On the one hand the heat capacity of the holes may increase or decrease, but on the other hand since the Gauss--Bonnet terms also affect their geometry (affecting $r_+$ and hence the thermodynamic volume $V$), their capacity to do work may also be affected one way or another. How the overall ratio between these two quantities  is affected is subtle, and we explore it in this paper.

\section{Einstein--Hilbert--Maxwell}
\label{sec:example}
Let's review and extend somewhat the computations for the  case of  heat engines made from Reissner--Nordstr\"om black hole solutions of the Einstein--Hilbert--Maxwell system, which has  bulk action in $D$--dimensions:
\begin{equation}
I=\frac{1}{16\pi }\int \! d^Dx \sqrt{-g} \left(R-2\Lambda -F^2\right)\ ,
\label{eq:action}
\end{equation}
where  the cosmological constant 
\begin{equation}
\Lambda=-\frac{(D-1)(D-2)}{2l^2}\ , 
\label{eq:cosmocon}
\end{equation}
 sets a length scale $l$. 
 
 \subsection{The Black Holes}
 The black hole has mass and charge  parameters $m$ and ${q}$, with metric\footnote{We're using the conventions of ref.\cite{Chamblin:1999tk}, and have chosen to work with  spacetimes  asymptotic to global AdS. }
\begin{equation}
ds^2 = -Y( r)dt^2
+ {dr^2\over Y(r)} + r^2 d\Omega^2_{D-2} 
\label{eq:staticform}
\end{equation}
where
\begin{equation}
Y( r) = 1-\frac{m}{r^{D-3}}+\frac{q^2}{r^{2D-6}}+\frac{r^2}{l^2}\ ,
\end{equation}
and $d\Omega^2_{D-2}$ is the metric on a round $D-2$ sphere with volume $\omega_{D-2}$, 
and  there is a gauge potential that is chosen to vanish on the horizon located at $r=r_+$, the largest positive real root of $Y(r)$:
\begin{equation}
A_t = \frac{q}{c}\left(\frac{1}{r_+^{D-3}}-\frac{1}{r^{D-3}}\right)\ , \quad {\rm with}\quad c=\sqrt{\frac{2(D-3)}{D-2}}\ .
\label{eq:gaugepotential}
\end{equation}
The mass and charge of the solution are given by:
\begin{equation}
M=\frac{(D-2)\omega_{D-2}}{16\pi} m \, \quad{\rm and}\quad Q=\sqrt{2(D-2)(D-3)}\left(\frac{\omega_{D-2}}{8\pi}\right) q \ .
\label{eq:paramters}
\end{equation}
\noindent
The  requirement of regularity of the Euclidean section fixes the temperature $T$ according to:
\begin{equation}
 T=\frac{1}{4\pi}Y^\prime \left.\right|_{r=r_+} = \frac{1}{4\pi}\left(16\pi p \frac{r_+}{(D-2)}+\frac{(D-3)}{r_+}-\frac{(D-3)q^2 }{r_+^{2D-5}}\right)\ ,
 \label{eq:teepee}
 \end{equation}
 where we have used that $p=-\Lambda/8\pi$ and equation~(\ref{eq:cosmocon}).
The entropy is $S=\omega_{D-2}r_+^{D-2}/4$. It is natural in writing $M$ as the enthalpy to express it in terms of the two variables, $p=-\Lambda/8\pi$ and~$S$. The statement that $r_+$ is the largest root of $Y(r)=0$ yields, after substituting for $p$: 
\begin{equation}
M(r_+,l)=\frac{(D-2)\omega_{D-2}}{16\pi}\left(r_+^{D-3}+\frac{q^2}{r_+^{D-3}}+16\pi p\frac{r_+^{D-1}}{(D-1)(D-2)}\right)\ ,
\label{eq:enthalpy}
\end{equation}
and $H(p,S)$ follows by substitution of $r_+$ in terms of $S$. For computing the thermodynamic volume $V=\partial H/\partial p|_S$ one may leave things in terms of $r_+$, getting:
\begin{equation}
V=\frac{\omega_{D-2}}{(D-1)} r_+^{D-1}\ .
\label{eq:volume}
\end{equation}
The temperature expression~(\ref{eq:teepee}) can be re--arranged into an equation of state (or rather a family of equations of state parameterized by $q$, which we will take to be fixed) in the $p{-}r_+$ plane, or equivalently (using equation~(\ref{eq:volume})) the $p{-}V$ plane:
\begin{equation}
p=\frac{(D-2)}{16\pi}\left(\frac{4\pi T}{r_+} -\frac{(D-3)}{r_+^2}+\frac{(D-3)q^2}{r_+^{2D-4}}\right)\ .
\end{equation}
%
Here, and more generally, it is worth noting the characteristic behaviour,
\begin{equation}
p V^{1/(D-1)}\sim T\ ,
\label{eq:idealgaslaw}
\end{equation}
  that dominates in the high temperature limit. In a sense, this is our analogue of the ideal gas limit for our black holes, giving  familiar hyperbolae in the $p{-}v$ plane where $v=V^{1/(D-1)}$. The structure at lower temperatures (giving a multivalued state curve, {\it etc}) results in non--trivial phase structure explored extensively in refs.\cite{Chamblin:1999tk,Chamblin:1999hg,Kubiznak:2012wp}. The details of the phase structure will not be of interest to us in this paper.
 
 \subsection{The Specific Heat}
 \label{sec:specificheat1}
 The most important quantity that we'll need to compute our engine efficiency is the specific heat  $T\partial S/\partial T$. This is computed  from our expression for temperature, 
 and we {\it could}  at this point substitute into it for $r_+$ in terms of $S$. Differentiation would then yield the specific heat. In preparation for what will come in our later examples, where substituting for $S$ is no longer elegant, we will follow the alternative route of leaving things written in terms of $r_+$ and instead differentiate  with respect to $T$, recovering   $\partial S/\partial T$ using the chain rule since the dependence $S(r_+)$ is known. The result  is, for general $D$:
 \begin{equation}
 C=T\frac{\partial S}{\partial T}=\left(1-\frac{4 r_+}{D-2}\frac{\partial p}{\partial T}\right)\left(\frac{\frac{16\pi }{(D-2)(D-3)} p\, r_+^{2D-4}+{r_+^{2D-6}}-{q^2 }{}}{\frac{16\pi }{(D-2)(D-3)} p\,r_+^{2D-4}-{r_+^{2D-6}}+{ (2D-5)q^2 }}\right)\frac{(D-2)\omega_{D-2}}{4}r_+^{D-2}\ .
 \label{eq:veryspecific}
 \end{equation}
 Since constant volume is also constant $r_+$, we see from equation~(\ref{eq:teepee}) that  $(\partial p/\partial T)_V=(D-2)/4r_+$. Hence,   the specific heat at constant volume vanishes $C_V=0$, while $C_p$ is given by setting $\partial p/\partial T=0$ in the expression. The vanishing of $C_V$ is the ``isochor equals adiabat'' result static black holes discussed above. For example, in $D=4$, with $\omega_2=4\pi$, we have\cite{Dolan:2010ha,Kubiznak:2012wp}, after substituting for $r_+$ in favour of $S$:
 \begin{equation}
 C_p = 2\pi r_+^{2}\left( {\frac { 8\pi pr_+^{4}+r_+^{2}-{q}^{2}  
 }{8\pi pr_+^{4}-r_+^{2}+3{q}^{2}}}\right)
 =2S\left(\frac{8pS^2+S-\pi q^2}{8pS^2-S+3\pi q^2}\right)\ ,
 \end{equation} 
in $D=5$, with $\omega_3=2\pi^2$:
\begin{equation}
C_p= \frac{3{\pi }^{2} }{2}r_+^{3}\left({\frac { 8\pi pr_+^{6}+3r_+^{4}-3{q}^{2}
  }{8\pi pr_+^{6}-3r_+^{4}+15{q}^{2}}}\right)
 = 3S \left( {\frac {32p{S}^{2}+6{(2\pi )}^{1/3}
 {S}^{4/3}-3{q}^{2}{\pi }^{3}  }{32p{S}^{2}-6{(2\pi)}^{1/3}{S}
^{4/3}+15{q}^{2}{\pi }^{3}}}\right)
 \ ,
\end{equation}
and in $D=6$, with $\omega_4=8\pi^2/3$:
\begin{equation}
C_p= \frac{8{\pi }^{2}}{3}r_+^{4}\left({\frac {  4\pi pr_+^{8}+3r_+^{6}-3{q}^{2}
 }{4\pi pr_+^{8}-3r_+^{6}+21{q}^{2}}}\right) 
= 4S\left({\frac {  12p{S}^{2}+3\cdot{6}^{1/2}{S}^{3/2}-4{q
}^{2}{\pi }^{3}  }{12p{S}^{2}-3 \cdot{6}^{1/2}{S}^{3/2}
+28{q}^{2}{\pi }^{3}}}\right)
 \ .
\end{equation}

\subsection{The Engine Efficiency}
Now we are ready to take the high temperature limit in order to get explicit expressions for the efficiency of the engine we designed above. With everything written in terms of $r_+$, all we need to do is solve for $r_+$ perturbatively in a large $T$ expansion, using equation~(\ref{eq:teepee}). From equations~(\ref{eq:teepee}),~(\ref{eq:veryspecific}), and~(\ref{eq:volume}), it is straightforward to extract the leading behaviour:
\begin{eqnarray}
&& r_+=\frac{(D-2)}{4}\frac{T}{p}+\cdots\nonumber\\
&& C_p=\left(\frac{(D-2)}{4}\right)^{D-1}\omega_{D-2}\left(\frac{T}{p}\right)^{D-2}+\cdots\nonumber\\
&& V=\frac{\omega_{D-2}}{D-1}\left(\frac{(D-2)}{4}\right)^{D-1}\left(\frac{T}{p}\right)^{D-1}+\cdots=\frac{1}{p}\int\! C_p dT+\cdots\ .
\label{eq:leading}
\end{eqnarray}
It is worth looking at some of the subleading results in specific dimensions. For $D=4$, we have (extending what was presented in ref.\cite{Johnson:2014yja}):
\begin{eqnarray}
&&r_+=\frac{1}{2}\frac{T}{p}-\frac{1}{4\pi T}+\frac18\frac{p(8\pi p q^2-1)}{\pi^2 T^3}+\cdots\ , \nonumber \\
&&V=\frac{4\pi}{3}r_+^3= \frac{\pi}{6p^3}T^3-\frac14 \frac{T}{p^2}+\frac{q^2}{T}+\cdots\ ,\nonumber \\
&& \int \!C_pdT=\frac{\pi}{6p^2}T^3+\frac18\frac{(16\pi pq^2-1)}{\pi T}+\cdots
\label{eq:4dexpansions}
\end{eqnarray} 
and in $D=5$ we have:
\begin{eqnarray}
&&r_+=\frac{3}{4}\frac{T}{p}-\frac{1}{2\pi T}-\frac{p}{3\pi^2 T^3}+\frac{4}{81}\frac{p^2(32q^2\pi^2p^2-9)}{\pi^3 T^5}+\cdots\ , \nonumber\\
&&V=\frac{\pi^2}{2}r_+^4= \frac{81}{512}\frac{\pi^2}{p^4}T^4-\frac{27}{64} \frac{\pi T^2}{p^3}+\frac{9}{64 p^2}+\frac43\frac{\pi p q^2}{T^2}+\cdots\ ,\nonumber \\ &&\int \!C_pdT=\frac{81}{512}\frac{\pi^2}{p^3}T^4-\frac{27}{128} \frac{\pi T^2}{p^2} +\frac{1}{96}\frac{(192\pi^2 p^2q^2-9)}{\pi T^2}+\cdots
\label{eq:5dexpansions}
\end{eqnarray} 
From equation~(\ref{eq:nicework}),  the efficiency is
\begin{equation}
\eta=\frac{W}{Q_H}=\left(1-\frac{p_4}{p_1}\right)\cdot\frac{p_1(V_2-V_1)}{Q_H}\ ,
\label{eq:efficiency}
\end{equation}
where $Q_H$ is given in equation~(\ref{eq:hothothot}). The volumes $V_2$ and $V_1$ are simply $V$ evaluated at $T_2$ and~$T_1$ respectively, with $p=p_1$ in each case. From equation~(\ref{eq:leading}) it is clear that the leading terms in our expressions for the volume and integrated specific heat will always give  unity at leading order for their ratio,  coming from the factor after the first parentheses. The subleading pieces are  of order $1/T^2$ and hence our  efficiency at large temperature is  
\begin{equation}
\eta=\left(1-\frac{p_4}{p_1}\right)+O\left(\frac{1}{T^2}\right) =\left(1-\frac{T_4}{T_1}\right)-O\left(\frac{1}{T^2}\right) =\eta^{\phantom{C}}_{\rm C}-O\left(\frac{1}{T^2}\right)  \ ,
\end{equation} where $\eta^{\phantom{C}}_{\rm C}=1-T_{\rm C}/T_H$, the Carnot efficiency of our engine, given its highest and lowest temperatures. This is an upper bound. (In the above, use was made of the ``ideal gas'' behaviour that is approached at high temperature, equation~(\ref{eq:idealgaslaw}),  the fact that corners $1$ and $4$ in our engine are at the same volume, and the fact that $\eta^{\phantom{C}}_{\rm C}$ is a maximum efficiency due to the Second Law.)
In each case, including the first subleading corrections, we have:
\begin{equation}
\eta= \left(1-\frac{p_4}{p_1}\right)\left(1-\frac{3}{2\pi}\frac{p_1}{(T_1^2+T_1T_2+T_2^2)}+\cdots\right)\ ,
\label{eq:efficiencyD4}
\end{equation}
for $D=4$, and 
\begin{equation}
\eta= \left(1-\frac{p_4}{p_1}\right)\left(1-\frac{4}{3\pi}\frac{p_1}{(T_1^2+T_2^2)}+\cdots\right)\ ,
\label{eq:efficiencyD5}
\end{equation}
for $D=5$.
We will evaluate $\eta$ further in section~\ref{sec:efficiency-alpha}, once we have computed the Gauss--Bonnet corrections.
\subsection{Remarks on  $q=0$}
It is interesting to note that at $q=0$ there is an exact relation between $r_+$ and $T$ that follows from the fact that the equation~(\ref{eq:teepee}) becomes a quadratic in $r_+$ in this case, for any $D$. The solution is:
\begin{equation}
r_+=\frac{(D-2)}{8p}\left(T\pm\sqrt{T^2-\frac{4p}{\pi }\frac{D-3}{D-2}}\right)\ .
\end{equation}
The plus sign choice gives the kind of solution we have seen already, with the large $T$ expansions above being seeded by  $r_+=[(D{-}2)/4]T/p+O(T^{-1})$. These are the famous ``large'' AdS black holes that grow with temperature\cite{Hawking:1982dh}. The negative sign has quite different behaviour, and as is clear from its large $T$ behaviour, $r_+=[(D{-}3)/4\pi] \cdot T^{-1}+O(T^{-2})$, corresponds to the famous ``small'' AdS black holes\cite{Hawking:1982dh}. It is not hard to see from equation~(\ref{eq:veryspecific}) that this behaviour immediately yields a strictly negative specific heat, which starts out as:
\begin{equation}
C_p=-\frac{D-2}{4}\omega_{D-2}\left(\frac{D-3}{4\pi}\right)^{D-2}\frac{1}{T^{D-2}}+O\left(\frac{1}{T^{D-4}}\right)\ ,
\end{equation}
and hence they are not obviously useful as heat engines.

\section{The Gauss--Bonnet Corrections}
Now we can turn to the main question of this paper, which is how the heat engines are affected by the presence  of a Gauss--Bonnet sector. Our action is:
\begin{equation}
I=\frac{1}{16\pi}\int\! d^Dx \sqrt{-g}\left[R-2\Lambda +\alpha_{\rm GB}(R_{\gamma\delta\mu\nu}R^{\gamma\delta\mu\nu}-4R_{\mu\nu}R^{\mu\nu}+R^2)- F^2\right]\ ,
\label{eq:gauss-bonnet-action}
\end{equation}
where we see that the Gauss--Bonnet parameter $\alpha_{\rm GB}$ has dimensions of ${\rm (length)}^2$. Since in  $D=4$ the  Gauss--Bonnet term is purely topological, we'll work in $D\ge5$ henceforth. 

\subsection{The Black Holes}
There is a charged static black hole solution of the form given in equations~(\ref{eq:staticform},\ref{eq:gaugepotential}) but now with\cite{Cai:2013qga}:
\begin{equation}
Y(r)=1+\frac{r^2}{2\alpha}\left(1-\sqrt{1+\frac{4\alpha m}{r^{D-1}}-\frac{4\alpha q^2}{r^{2D-4}}-\frac{4\alpha}{l^2}}\right) \ ,
\label{eq:why}
\end{equation}
where $\alpha=(D-3)(D-4)\alpha_{\rm GB}$. The parameters $M$ and $q$ set the mass and charge as before\footnote{We are using the conventions of ref.\cite{Cai:2013qga}, with a slight modification of the Maxwell sector.} (see equation~(\ref{eq:cosmocon})), and again the cosmological constant  is set by $l$ according to equation~(\ref{eq:paramters}).

Notice that  since the $m=q=0$ case, defining the vacuum solution, ought to be well--defined ({\it i.e.,} neither imaginary nor nakedly singular),  for a given value of $l$ (and hence $\Lambda$) $\alpha$ cannot be arbitrary\cite{Boulware:1985wk}, but in fact must be constrained by $0\leq {4\alpha}/{l^2}\leq 1$. For later use we can write this in terms of the pressure as:
\begin{equation}
  0\leq \alpha \leq \alpha_*\ , \quad {\rm where}\quad  \alpha_*={(D-1)(D-2)}/{64\pi p} \ .
\label{eq:constrain-alpha}
\end{equation}
While the above interpretation is the standard one used in the literature, one could regard this  in a different light, treating $\alpha$ as an arbitrary parameter setting the strength of the Gauss--Bonnet term,  and then thinking of equation~(\ref{eq:constrain-alpha}) as setting a maximum value that the pressure, $p$,  can attain ({\it i.e.,} equivalent to a maximum value of $|\Lambda|$ or a minimum value for the scale~$l$). While this is interesting, we will not take this approach, given our motivations stated in the Introduction: We are considering the Gauss--Bonnet term as a correction to the Einstein--Maxwell system and the heat engines defined therein. So we will be defining a specific heat engine ---which involves specifying a range of pressures, $p$, over which it operates--- and then seeing the effects of turning on the Gauss--Bonnet term (controlled by $\alpha$) on the physics. Therefore we will consider only the physical range of $\alpha$ that is consistent with those defining  pressures.

 The  horizon radius $r_+$ of the black hole is set by the largest  root of $Y(r_+)=0$, which gives us an equation for  $M$, generalizing equation~(\ref{eq:enthalpy}). 
\begin{equation}
M=\frac{(D-2)\omega_{D-2}}{16\pi}\left(\alpha r_+^{D-5}+r_+^{D-3}+\frac{q^2}{r_+^{D-3}}+16\pi p \frac{r_+^{D-1}}{(D-1)(D-2)}\right)\ ,
\end{equation}
where we have replaced $l$ by $p$ using $p=-\Lambda/8\pi$ and equation~(\ref{eq:cosmocon}). This and the next few steps reproduces the results in ref.\cite{Cai:2013qga}, but with our conventions for the Maxwell sector.
The temperature comes from the first derivative of $Y$ at the horizon, in the usual way:
\begin{equation}
T=\frac{Y^\prime(r_+)}{4\pi}=\frac{1}{4\pi r_+(r_+^2+2\alpha)}\left(\frac{16\pi p r_+^4}{(D-2)}+(D-3)r_+^2+(D-5)\alpha-(D-3)\frac{q^2}{r_+^{2D-8}}\right)\ ,
\label{eq:GBeqnofstate}
\end{equation}
The function $M$ defines our enthalpy $H(p,S)$, from which the entropy  can be computed as:
\begin{equation}
S=\int_0^{r_+}\frac{1}{T}\left.\frac{\partial M}{\partial r}\right|_{q,p} dr=\frac{\omega_{D-2}}{4} r_+^{D-2}\left(1+\frac{2(D-2)}{(D-4)}\frac{\alpha}{r_+^2}\right)\ .
\end{equation}
Since the pressure term in $M$ is unaffected (and all the $r_+$ dependence determines the $S$ dependence) it is clear that the thermodynamic volume again turns out to be that given in equation~(\ref{eq:volume}). 

Again we see that for a given $\alpha$, $S$ and $V$ are not independent, and so the structure of what we saw for $\alpha=0$ will follow again.  We have that $C_V=0$ and we can compute $C_p$, so for constructing heat engines using our black holes, the rectangular engine cycle of the last section (see figure~\ref{fig:cyclesb} and surrounding discussion) is still extremely natural since again. We are now ready to  compare the efficiency of our new engines at $\alpha\neq 0 $ to that of the last section ($\alpha=0$).

\subsection{The Specific Heat}
Using the procedures of section~\ref{sec:specificheat1}, we can compute the specific heat in terms of $r_+$ rather straightforwardly. The full $D$--dependent expression is rather messy, and so we present it as follows. Writing the temperature in equation~(\ref{eq:teepee}) as $T=RU$ where $R^{-1}=4\pi r_+(r_+^2+2\alpha)$ and 
\begin{equation}
U=\frac{16\pi p r_+^4}{(D-2)}+(D-3)r_+^3+(D-5)\alpha-(D-3)\frac{q^2}{r_+^{2D-8}} \ ,
\end{equation}
 we get:
\begin{equation}
C_p=(D-2)\frac{\omega_{D-2}}{4}\frac{U(r_+^2+2\alpha)[r_+^{D-3}+2(D-2)\alpha r_+^{D-5}]}{U^\prime|_p(r_+^2+2\alpha)-U(r_+^5+2\alpha r_+^3+2r_+)}\ .
\end{equation}
In $D=5$  the result is:
\begin{equation}
C_p=\frac{3\pi^2}{2}\frac { \left( 8\pi pr_+^{6}+3r_+^{4}-3q^{2} \right)  \left( r_+^{2}+2\alpha \right)^{2}r_+}{(8r_+^{8}\pi p-3
r_+^{6}+15r_+^{2}{q}^{2}+48\alpha\pi p\,r_+^{6}+6\alpha r_+
^{4}+18\alpha{q}^{2})} \ .
\end{equation}
while in $D=6$   it is:
\begin{equation}
C_p=\frac{8{\pi }^{2}}{3}\,{\frac { \left( 4\pi pr_+^{8}+3r_+^{6}+\alpha r_+^{4}-3{
q}^{2} \right)  \left( r_+^{2}+2\alpha \right)^{2}r_+^{2}
}{(4r_+^{10}\pi p-3r_+^{8}+3r_+^{6}\alpha+21r_+^{2}q^{2}+24
\alpha\pi pr_+^{8}-2{\alpha}^{2}r_+^{4}+30\alpha{q}^{2})}}
\end{equation}

\subsection{The Engine Efficiency}
\label{sec:efficiency-alpha}
Again, working in the  high temperature limit in order to be able to extract expressions for the integrated specific heat and the volume, we solve equation~(\ref{eq:GBeqnofstate}) iteratively for the horizon  $r_+$ as a function of $(T, p, q)$, giving, for $D=5$:
\begin{eqnarray}
&&r_+= \frac{3}{4}{\frac {T}{p}}+\frac16{\frac {(16\pi \alpha p-3)}{\pi T}}-\frac{1}{27}
{\frac {p \left( 32\pi \alpha p-3 \right)  \left( 16\pi 
\alpha p-3 \right) }{{\pi }^{2}{T}^{3}}}\nonumber\\&&\hskip3.4cm+{\frac {4}{243}}{\frac {{p
}^{2} \left( -27+720 \pi  \alpha p+14336 {\pi }^{3}{\alpha}^{3}{p}
^{3}+96 {q}^{2}{\pi }^{2}{p}^{2}-5760 {\pi }^{2}{\alpha}^{2}{p}^{2}
 \right) }{{\pi }^{3}{T}^{5}}}+\cdots
\nonumber\\
&&V= {\frac {81}{512}} {\frac {{\pi }^{2}{T}^{4}}{{p}^{4}}}+{\frac {9}{64}
} {\frac {\pi   \left( 16 \pi  \alpha p-3 \right) {T}^{2}}{{p}^{3
}}}-{\frac {1}{64}} {\frac {256 {\pi }^{2}{\alpha}^{2}{p}^{2}-9}{{p}
^{2}}}\nonumber\\&&\hskip5cm+\frac19 {\frac {(9 \alpha+512 {\pi }^{2}{p}^{2}{\alpha}^{3}+12 
\pi  p{q}^{2}-144 \pi  p{\alpha}^{2})}{{T}^{2}}} +\cdots
\nonumber\\
&&\int\! C_p dT=  {\frac {81}{512}} {\frac {{\pi }^{2}{T}^{4}}{{p}^{3}}}+{\frac {9}{128
}} {\frac {\pi   \left( 32 \pi  \alpha p-3 \right) {T}^{2}}{{p}^{
2}}}\nonumber\\&&\hskip2.5cm+{\frac {1}{288}} {\frac {(-27+864 \pi  \alpha p+16384 {\pi }^
{3}{\alpha}^{3}{p}^{3}+576 {q}^{2}{\pi }^{2}{p}^{2}-6912 {\pi }^{2}{
\alpha}^{2}{p}^{2})}{\pi  {T}^{2}}}+\cdots
\  ,
\label{eq:5dexpansions-alpha}
\end{eqnarray}
a deformation, by $\alpha$, of our expressions in equation~(\ref{eq:5dexpansions}). For $D=6$ we have:
\begin{eqnarray}
&&r_+=  
{\frac {T}{p}}+\frac{1}{4} {\frac {(8 \pi  \alpha p-3)}{\pi  T}}-\frac{1}{16} {
\frac {p \left( -68 \pi  \alpha p+128 {\pi }^{2}{\alpha}^{2}{p}^{2
}+9 \right) }{{\pi }^{2}{T}^{3}}}\nonumber\\
&&\hskip5cm+\frac{1}{32} {\frac {{p}^{2} \left( -1360 
{\pi }^{2}{\alpha}^{2}{p}^{2}+336 \pi  \alpha p-27+1792 {\pi }^{3}
{\alpha}^{3}{p}^{3} \right) }{{\pi }^{3}{T}^{5}}}+\cdots
\nonumber\\
&&V= 
{\frac {8}{15}} {\frac {{\pi }^{2}{T}^{5}}{{p}^{5}}}+\frac{2}{3} {\frac {
\pi   \left( 8 \pi  \alpha p -3\right) {T}^{3}}{{p}^{4}}}-\frac{1}{6} {
\frac { \left( 28 \pi  \alpha p-9 \right) T}{{p}^{3}}}+\frac{4}{3} {\frac 
{\pi  {\alpha}^{2} \left( -5+16 \pi  \alpha p \right) }{T}}\nonumber\\
&&\hskip4cm -\frac{1}{24} 
{\frac {(-48 {\pi }^{2}{p}^{2}{q}^{2}+724 \pi  p{\alpha}^{2}-3456 {
\pi }^{2}{p}^{2}{\alpha}^{3}-45 \alpha+5120 {p}^{3}{\alpha}^{4}{\pi 
}^{3})}{\pi  {T}^{3}}}+\cdots
 \nonumber\\
&&\int\! C_p dT=   
{\frac {8}{15}} {\frac {{\pi }^{2}{T}^{5}}{{p}^{4}}}+\frac{4}{3} {\frac {
\pi   \left( 4 \pi  \alpha p -1\right) {T}^{3}}{{p}^{3}}}-
 \left( {\frac {40}{3}} \pi  p{\alpha}^{2}-{\frac {64}{3}} {\pi }^{
2}{p}^{2}{\alpha}^{3}-2 \alpha \right) \frac{1}{T}     \label{eq:6dexpansions-alpha} \\  
&&\hskip1cm-{\frac {1}{96}} {
\frac {(-256 {q}^{2}{\pi }^{3}{p}^{3}-18432 {\pi }^{3}{\alpha}^{3}{p}
^{3}+5792 {\pi }^{2}{\alpha}^{2}{p}^{2}+20480 {\alpha}^{4}{p}^{4}{
\pi }^{4}-720 \pi  \alpha p+27)}{{\pi }^{2}{T}^{3}}}+\cdots\nonumber
\end{eqnarray}

The efficiency (at high $T$) of our engine cycle in figure~\ref{fig:cyclesb} is given (perturbatively) by inserting these new quantities into equation~(\ref{eq:efficiency}), and we can now answer the question of  how our Gauss--Bonnet deformation has affected --increased or decreased-- the efficiency of our engine. We can follow the efficiency as a function of~$\alpha$, seeing how it behaves as we move away from $\alpha=0$. From the form of the expansion, it is also clear that we must  be careful, at any given order in the $1/T$ expansion, to not take $\alpha$ (or really, the product $\alpha p$) too large so as not to compromise the accuracy of the expansion. 
Before we do that exploration, we must decide what parameters of the cycle we specify and  hold fixed as we compare different engines by changing $\alpha$. There is a variety of choices, but we'll study two schemes in what follows.


\subsubsection{Scheme 1}
Here, for our engine cycle (see figure~\ref{fig:cyclesb}) we  specify the two operating pressures  $(p_1,p_4)$ and the two temperatures $(T_1,T_2)$. Specifying those two temperatures is in some sense in natural accordance with how the input heat $Q_H$ is defined in equation~(\ref{eq:hothothot}), using  the pressure $p_1$ of the isobar and the beginning and ending temperatures of the isobaric expansion phase. We can compute the efficiency in this scheme as a function of $\alpha$, seeing how it moves away from the standard set by~$\alpha=0$.


Actually, at a given value of $\alpha$ another important  standard to compare to is the Carnot efficiency $\eta^{\phantom{C}}_{\rm C}=1-T_{\rm C}/T_H$ which is the efficiency we'd get with a reversible heat engine operating between those same two temperatures. (Recall that $T_{\rm C}$ and $T_H$ are the lowest and highest temperatures our engine can attain.) Note however that although we've specified $T_H\equiv T_2$, $\eta^{\phantom{C}}_{\rm C}$ changes with  $\alpha$ since $T_{\rm C}$ does: The equation of state~(\ref{eq:GBeqnofstate}) must be used to determine $T_4\equiv T_{\rm C}$. In fact, as~$\alpha$ increases $T_{\rm C}$ falls, as can be seen by using\footnote{We thank Shao--Wen Wei for pointing  out an error in earlier versions of this manuscript made while using  perturbative methods for determining $T_{\rm C}$ which break down for small enough temperature.} the equation of state~(\ref{eq:GBeqnofstate}), meaning that the Carnot cycle available is {\it more} efficient with increasing $\alpha$.  (We will return to this point quantitatively below.) 

Now, specifying  a  particular choice of parameters of the heat engine's cycle includes picking the pressures $p_1$ and $p_4$. Once that is done, the actual values that $\alpha$  can take will be bounded above by $\alpha_*$ in equation~(\ref{eq:constrain-alpha}), coming from the constraint to have a physical vacuum solution.  We  insert $p=p_1$ into that equation since that is the largest pressure the engine will encounter in its cycle, and therefore gives the more strict bound.

\begin{figure}[h]
{\centering
\subfigure[]{\includegraphics[width=2.5in]{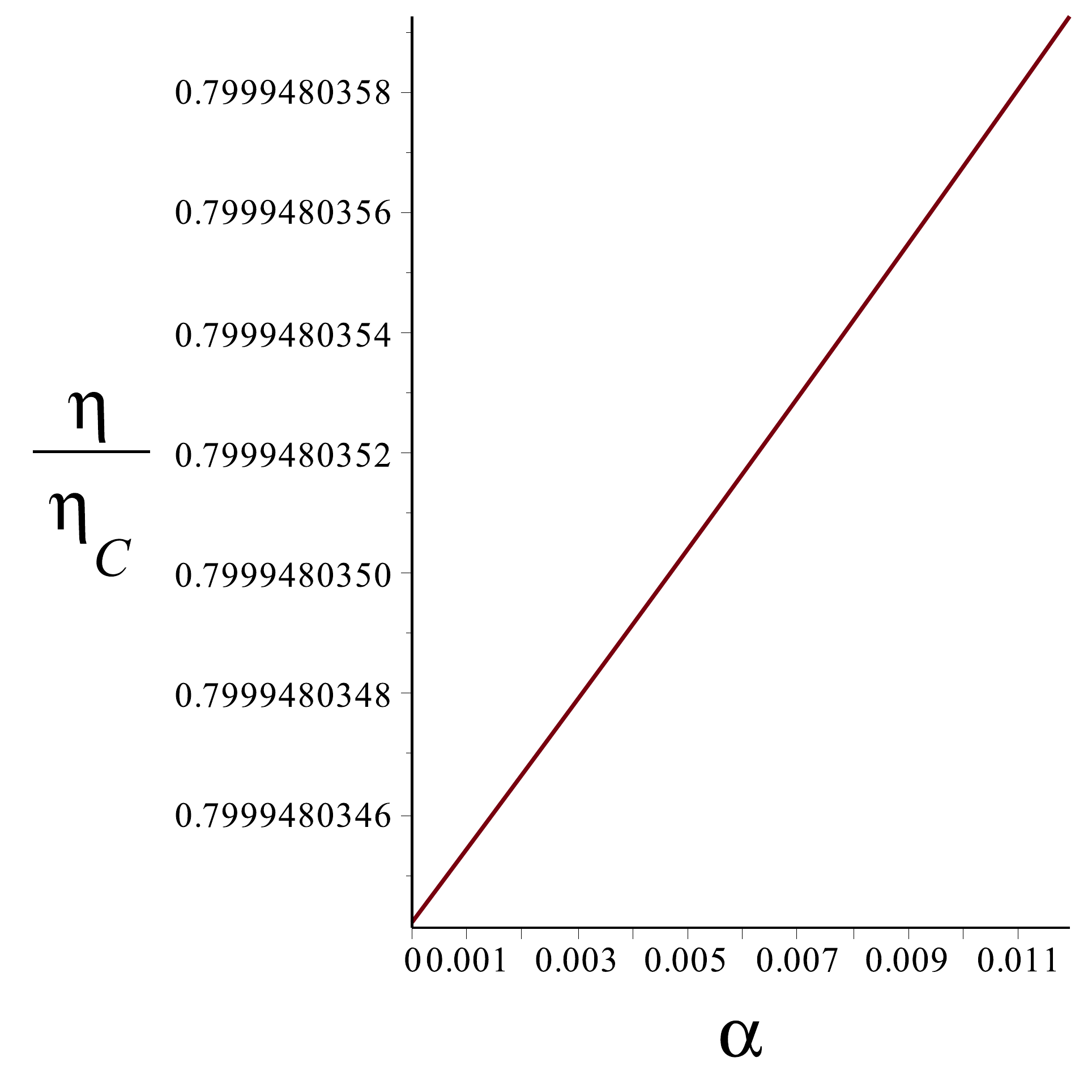} }\hspace{1.9cm}
\subfigure[]{\includegraphics[width=2.5in]{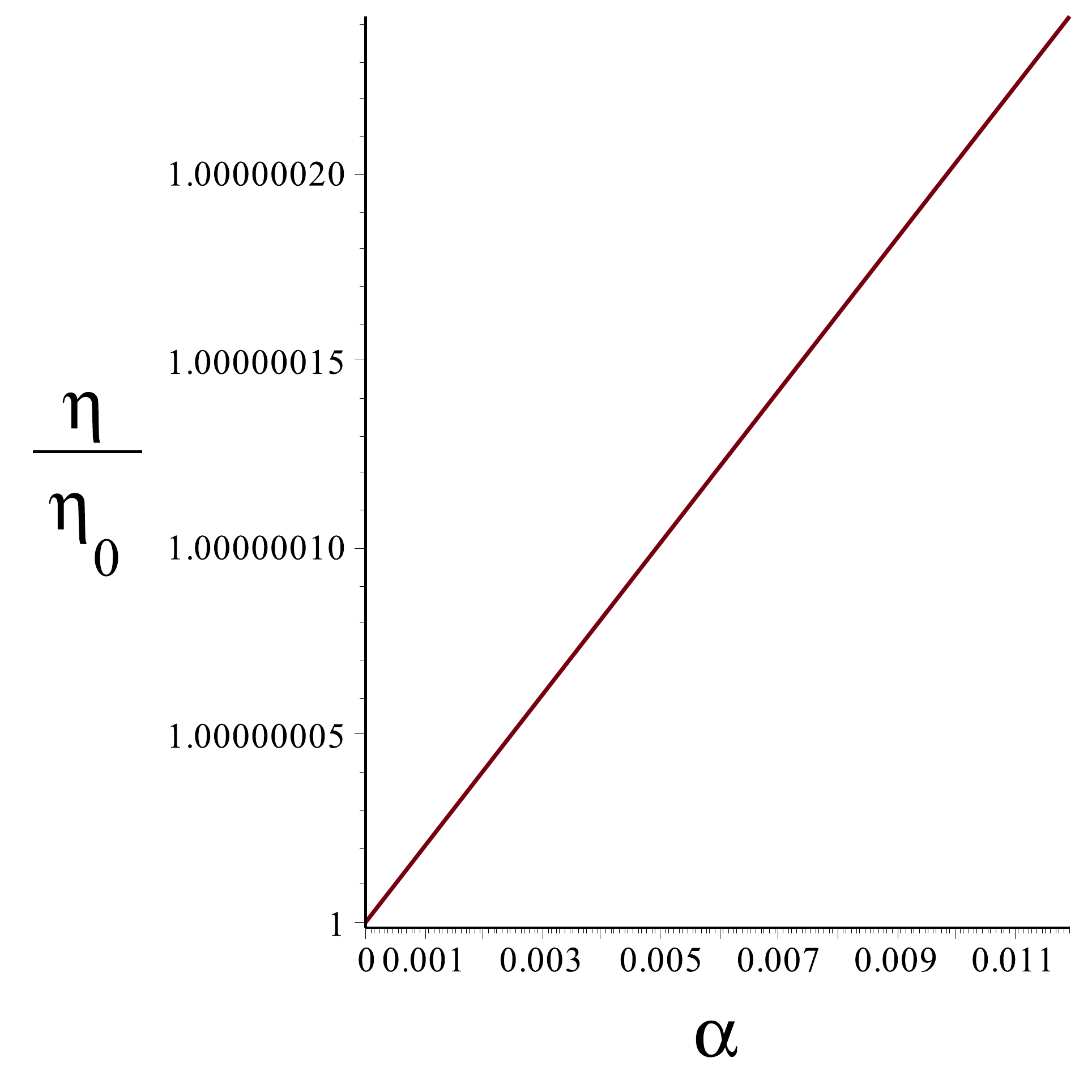} }

   \caption{\footnotesize  (a) The ratio $\eta/\eta_{\rm C}$ {\it vs} $\alpha$ in scheme~1,  plotted over the  physical range of $\alpha$ determined  by equation~(\ref{eq:constrain-alpha}). (b)  The ratio $\eta/\eta_0$ over that same  range. This is for $D=5$.  (Here, we've chosen the values $p_1=5, p_4=3, T_1=50, T_2=60,$ and $q=0.1$.)}   \label{fig:efficiency-compare-2}
}
\end{figure}

With those subtleties in mind, we can now study the efficiency of our engine at various physical values of $\alpha$, and also compare to the available Carnot efficiency at the same $\alpha$.  
As an example, in figure~\ref{fig:efficiency-compare-2} the efficiencies are compared  over the physical range $0<\alpha<\alpha_*$, for a particular choice of parameters. 

Figure~\ref{fig:efficiency-compare-2}(a) shows the ratio $\eta/\eta^{\phantom{C}}_{\rm C}$, and it  increases over the physical range, for $D=5$. Defining $\eta^{\phantom{9}}_0$ as the efficiency at $\alpha=0$, figure~\ref{fig:efficiency-compare-2}(b) shows that the ratio $\eta/\eta^{\phantom{9}}_0$  increases over the physical range, again for $D=5$. Exploring different choices of parameters reveals that the increase behaviour in figure~\ref{fig:efficiency-compare-2}(b) is robust, while that of figure~\ref{fig:efficiency-compare-2}(a) is not: sometimes it can be a decrease. At least in this  large $T$ regime, we can understand this from the expressions in equation~(\ref{eq:5dexpansions-alpha}). It is sufficient to truncate the expressions for the volume and the integrated specific heat to keep the leading positive powers of $T$ in the expansion.  Then, inserting them into the expression~(\ref{eq:efficiency}) for the efficiency yields
\begin{equation}
\eta=\left(1-\frac{p_4}{p_1}\right)(1+A_1 x)(1+A_2 x)^{-1}\ ,
\label{eq:efficiencytruncated}
\end{equation}
where
\begin{equation}
A_1=\frac{8}{9\pi} p_1(16\pi\alpha p_1 - 3)\ ,\quad A_2=\frac{4}{9\pi} p_1(32\pi\alpha p_1 - 3)\ ,\quad {\rm and }\quad  x =\frac{T_2^2-T_1^2}{T_2^4 - T_1^4}\ .
\label{eq:coeffsD5}
\end{equation}
Expanding in small $x$ gives the ratio:
\begin{equation}
\frac{\eta}{\eta_0}=1 +\frac{512}{27}\frac{p_1^3}{\pi} x^2 \alpha+\cdots \ ,
\end{equation}
explaining the robustness of the results of the type shown in figure~\ref{fig:efficiency-compare-2}(b). That the result (increase or decrease with $\alpha$) for $\eta/\eta^{\phantom{C}}_{\rm C}$  is parameter dependent follows from the fact that the Carnot efficiency requires a determination of $T_{\rm C}=T_4$ through the equation of state, picking up dependence on the way scheme~1 is defined through parameters specified at other corners of the cycle. Some algebra  shows that at leading  order in $\alpha$:
\begin{equation}
T_{\rm C} = \frac{p_4}{p_1} T_1 + \frac{2}{3\pi} (p_1-p_4)\frac{1}{T_1}
-{\frac {2}{27{\pi }^{2}}}\,{\frac {p_1}{p_4} \left( p_1-p_4 \right) 
 \left( -3p_1+p_4(64p_1\pi \alpha-9)
 \right) }\frac{1}{T_1^3}
+O\left(\frac{1}{T_1^5}\right)\ ,
\label{eq:coldtempD6}
\end{equation}
and so since $T_{\rm C}$ is falling linearly with $\alpha$, the Carnot efficiency $\eta^{\phantom{C}}_{\rm C}=1-T_C/T_2$  rises at that rate. (The first two terms of the large $T$ expansion~(\ref{eq:5dexpansions-alpha}) for $V$ evaluated at different corners of the cycle was the  approach used to obtain this.) The key issue then is how fast $\eta^{\phantom{C}}_{\rm C}$ rises with $\alpha$  compared to $\eta$. The coefficient is $\sim p_1^2(p_1-p_4)T_1^{-3}T_2^{-1}$, while the rise of $\eta$  is governed by  coefficient $\sim  p_1^3 x^2$ (recall that $x$ is a combination of $T_1$ and $T_2$ given in equation~(\ref{eq:coeffsD5}), and  we have ignored pure numbers that don't matter in the argument). Therefore for an appropriate choice of parameters $(T_1,T_2,p_1,p_4)$ of the engine,  $\eta^{\phantom{C}}_{\rm C}$ can  increase with~$\alpha$ more slowly than~$\eta$, giving a positive slope  for $\eta/\eta^{\phantom{C}}_{\rm C}$. Different choices can instead make  $\eta^{\phantom{C}}_{\rm C}$ increase  with $\alpha$ more quickly than $\eta$, making the slope of $\eta/\eta^{\phantom{C}}_{\rm C}$ negative.

\begin{figure}[h]
{\centering
\subfigure[]{\includegraphics[width=2.5in]{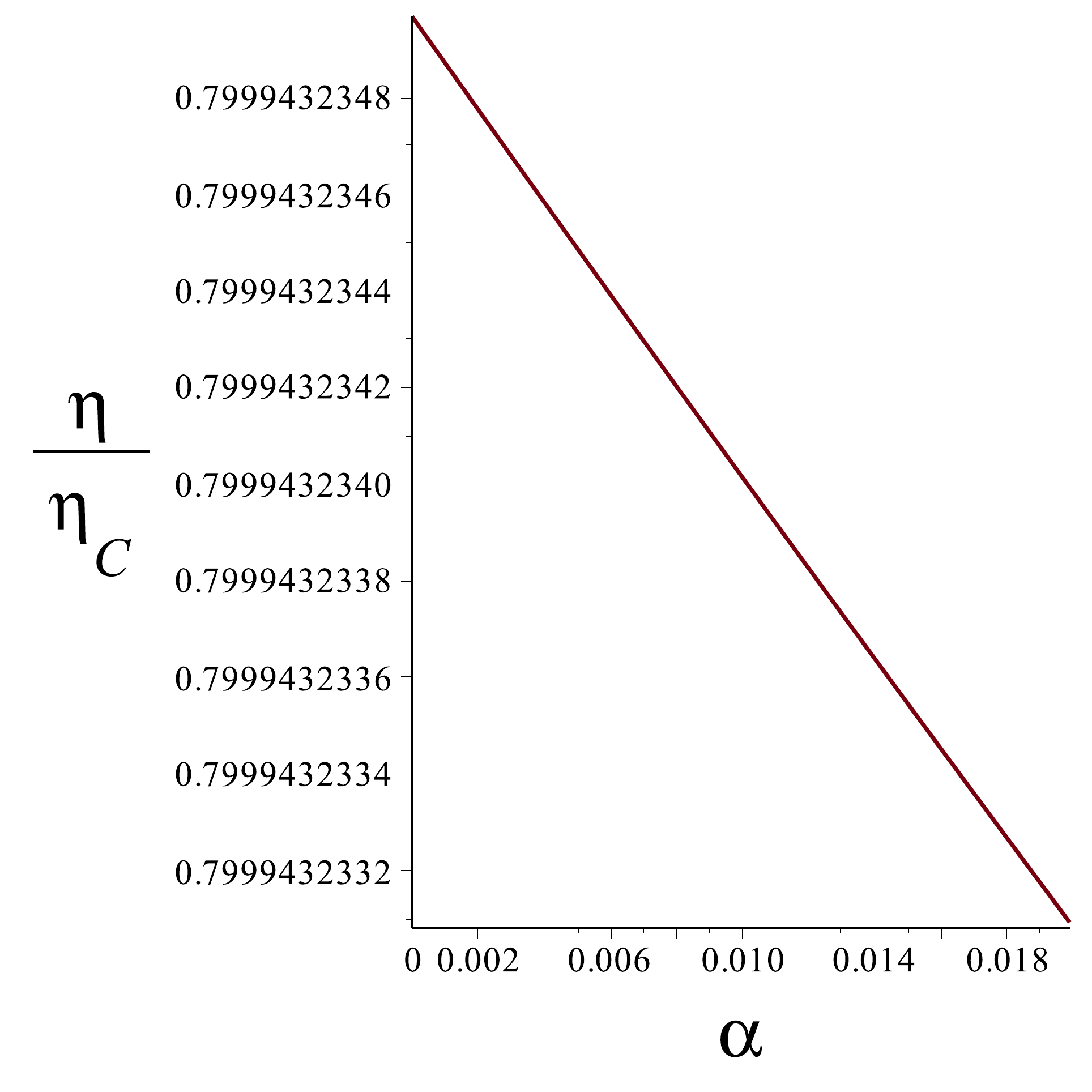} }\hspace{1.9cm}
\subfigure[]{\includegraphics[width=2.5in]{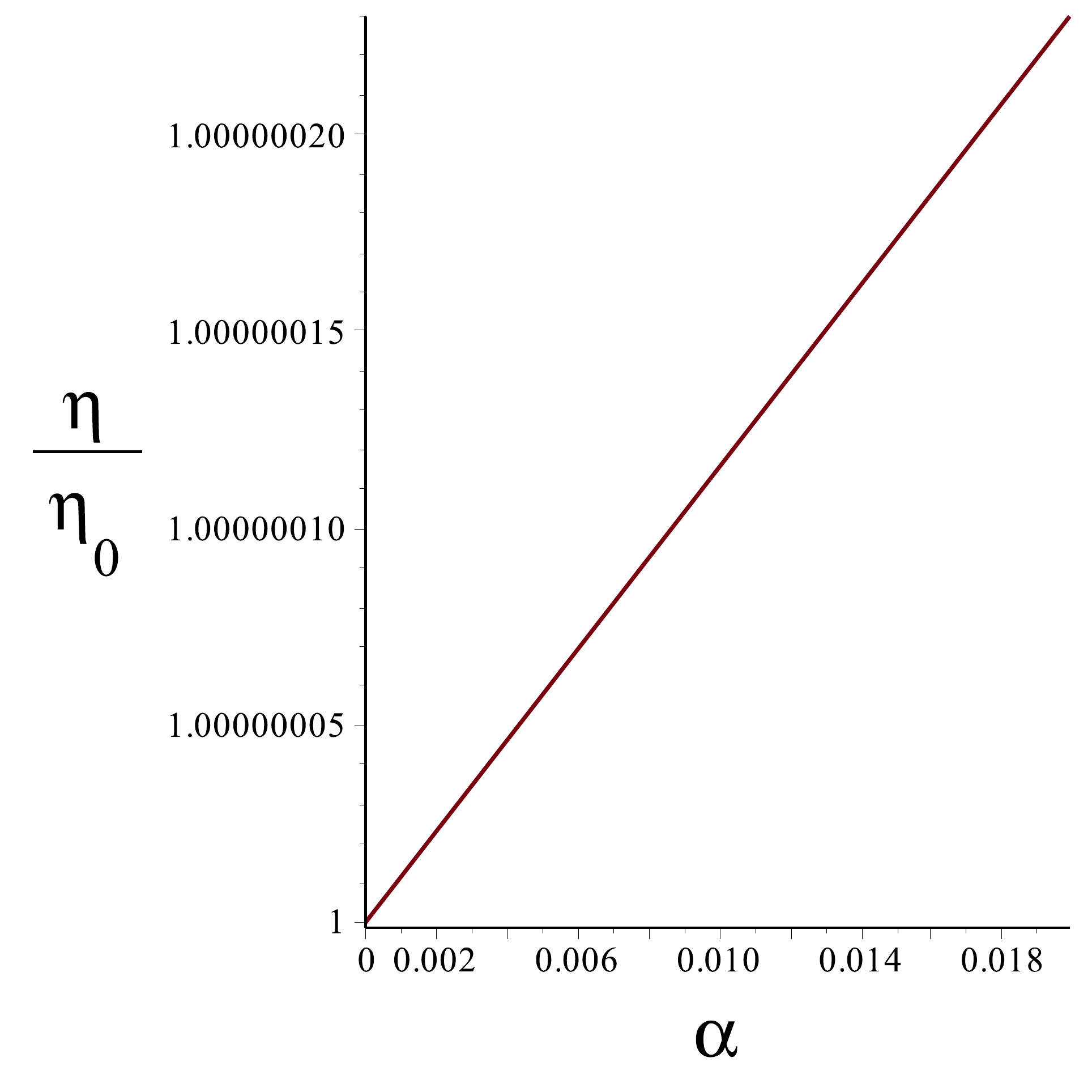} }

   \caption{\footnotesize  (a) The ratio $\eta/\eta_{\rm C}$ {\it vs} $\alpha$ in scheme~1,  plotted over the  physical range of $\alpha$ determined  by equation~(\ref{eq:constrain-alpha}). (b)  The ratio $\eta/\eta_0$ over that same  range.  This is for $D=6$. (Here, we've chosen the same values  as for $D=5$: $p_1=5, p_4=3, T_1=50, T_2=60,$ and $q=0.1$.)}   \label{fig:efficiency-compare-2x}
}
\end{figure}

Moving to one dimension higher, figures~\ref{fig:efficiency-compare-2x}(a) and~\ref{fig:efficiency-compare-2x}(b) show sample behaviour for $D=6$ for the same defining values of the state variables, which  by contrast has a decrease for $\eta/\eta^{\phantom{C}}_{\rm C}$, and an increase for $\eta/\eta_0$. Again, exploration shows that the behaviour of $\eta/\eta_0$  is apparently robust, while~$\eta/\eta^{\phantom{C}}_{\rm C}$ is  parameter dependent. This is backed up again by a study of the large $T$ expressions in equation~(\ref{eq:6dexpansions-alpha}).  Curiously, the volume expression has one extra positive power of $T$ (a linear one) while the  integrated specific heat does not. Ignoring the linear term at first, analysis very similar to above yields an expression for the efficiency of the same form as in equation~(\ref{eq:efficiencytruncated}) but now with:
\begin{equation}
A_1=\frac{5}{4\pi} p_1(8\pi\alpha p_1 - 3)\ ,\quad A_2=\frac{5}{2\pi} p_1(4\pi\alpha p_1 - 1)\ ,\quad {\rm and }\quad  x =\frac{T_2^3-T_1^3}{T_2^5 - T_1^5}\ .
\end{equation}
Expanding again in small $x$ gives the ratio:
\begin{equation}
\frac{\eta}{\eta_0}=1 +\frac{25}{2}\frac{p_1^3}{\pi} x^2 \alpha+\cdots \ ,
\end{equation}
which is encouragingly positive. The linear term in $T$, which has a negative sign, must be taken into account. Further algebra shows that its contribution at this order gives $(-35/4\pi) p_1^3 y$ where $y=(T_2-T_1)/(T_2^5 - T_1^5)$. This is not enough to change the sign, and so indeed the increase in $\eta/\eta_0$ is robust. Again, that the sign of the slope of   $\eta/\eta^{\phantom{C}}_{\rm C}$ is parameter dependent follows from the equation of state dependence of $T_{\rm C}$. In this case
\begin{equation}
T_{\rm C} = \frac{p_4}{p_1} T_1 + \frac{3}{4\pi} (p_1-p_4)\frac{1}{T_1}
-{\frac {9}{16{\pi }^{2}}}\,{\frac {p_1}{p_4} \left( p_1-p_4 \right) 
 \left( -p_1+p_4(8p_1\pi \alpha-2)
 \right) }\frac{1}{T_1^3}
+O\left(\frac{1}{T_1^5}\right)\ ,
\label{eq:coldtempD6}
\end{equation}
{\it i.e.,} of similar form as for $D=5$ with slightly different coefficients (of the same signs as before), and so the analysis goes through very  similarly.

 We now turn to scheme 2.

\subsubsection{Scheme 2}

\begin{figure}[h]
{\centering
\subfigure[]{\includegraphics[width=2.8in]{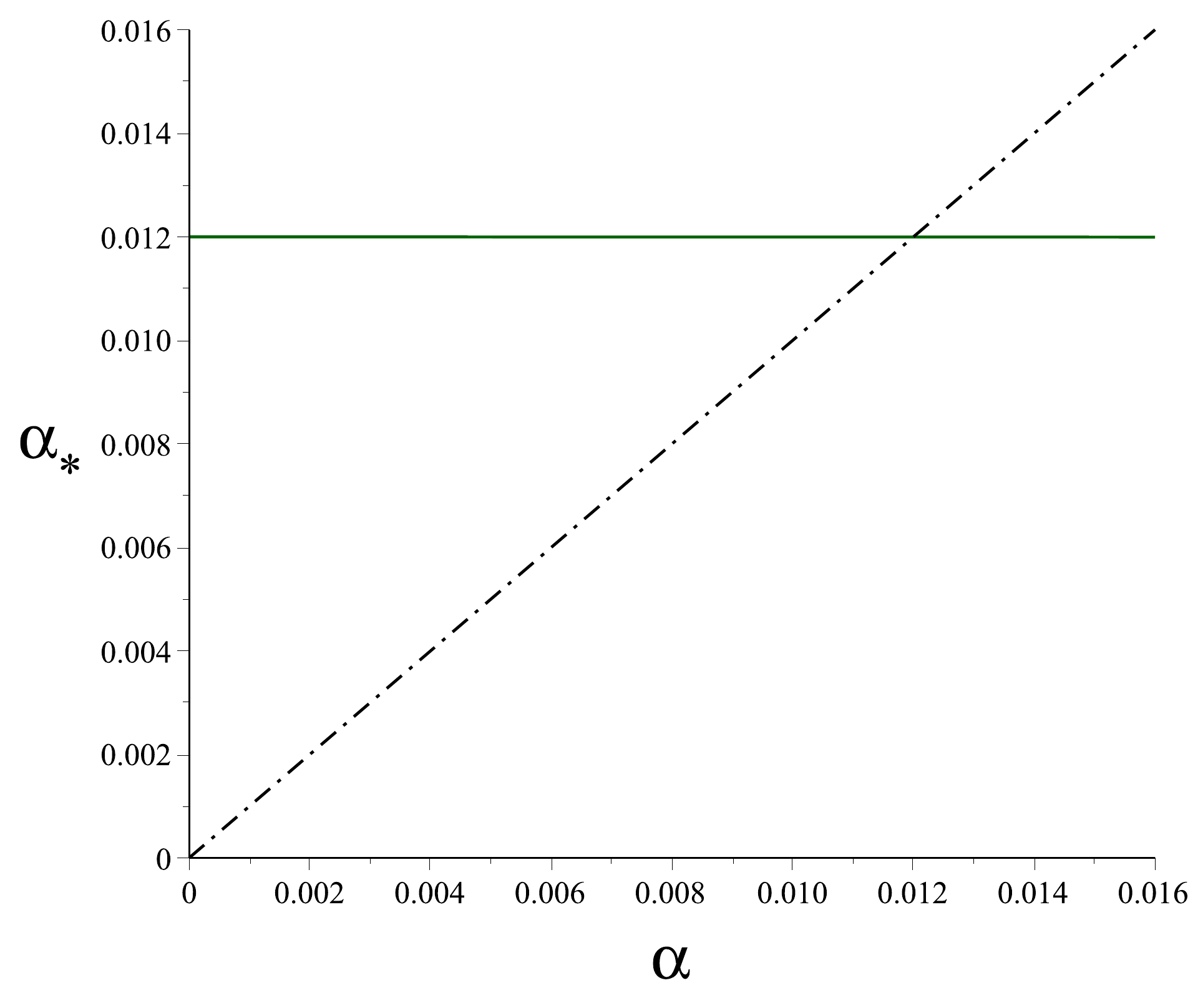}}\hspace{1.5cm}
\subfigure[]{\includegraphics[width=2.8in]{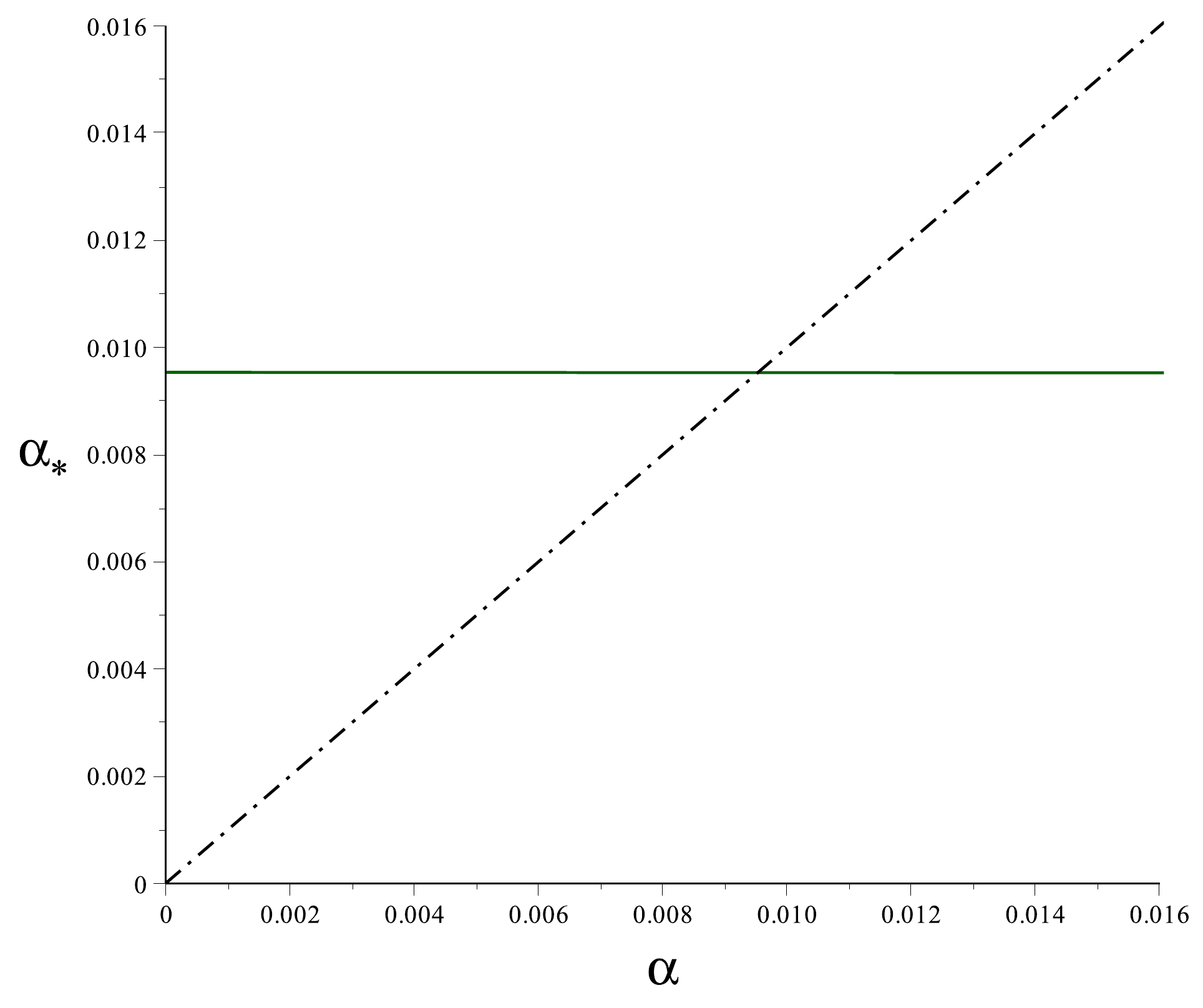} }
   \caption{\footnotesize  
   The determination of the  physical range of $\alpha$ according to equation~(\ref{eq:constrain-alpha}), in scheme~2, taking into account that the pressures of the cycle are $\alpha$--dependent in that scheme. (See text.)  Case (a) is for $D=5$ and case (b) is for $D=6$. (For both cases, we chose $T_2\equiv T_H=60$, $V_2=33000$, $T_4\equiv T_{\rm C}=30$, $V_4=15500$, $q=0.1$, and for $D=5$ the highest allowed $\alpha$ was approximately 0.011993 while for $D=6$ it was approximately 0.009529.)}   \label{fig:efficiency-compare-3}
}
\end{figure}

In this scheme for our engine (referring again to figure~\ref{fig:cyclesb}) we instead specify the temperatures $(T_2,T_4)$, equivalent to specifying $(T_H,T_{\rm C})$, as well as the volumes $(V_2,V_4)$ (which also gives the pair $(V_3,V_1)$). This might be considered a natural choice if one  is constructing an engine with specific initial and final volumes in mind. Now the Carnot efficiency  $\eta^{\phantom{C}}_{\rm C}$ is fixed for all $\alpha$. Instead, however, the pressures $p_1=p_2$ and $p_4=p_3$ must be determined using the equation of state, and so are now $\alpha$--dependent. 
The upper bound on $\alpha$ coming from equation~(\ref{eq:constrain-alpha}) is more subtle to implement here since the pressures in our engine are $\alpha$ dependent. However, one can proceed as follows. At a given value of $\alpha$, we can compute the largest pressure ($p_1=p_2$) that occurs in the engine, and check whether the limiting $\alpha_*$ obtained from that $p_1$ is less than~$\alpha$. If it is, then that particular engine is not physical and we have strayed beyond the maximum $\alpha$. Put differently, as a function of $\alpha$ there is a curve $\alpha_*(\alpha)=(D-1)(D-2)/64\pi p_1(\alpha)$ which starts out at some positive value at~$\alpha=0$. At some point it meets the line of slope unity ($\alpha_*=\alpha$), and that crossing point determines $\alpha$'s maximum value. See figure~\ref{fig:efficiency-compare-3}(a) and (b), for the case $D=5$ and $D=6$ respectively. 

\begin{figure}[h]
{\centering
\subfigure[]{\includegraphics[width=2.5in]{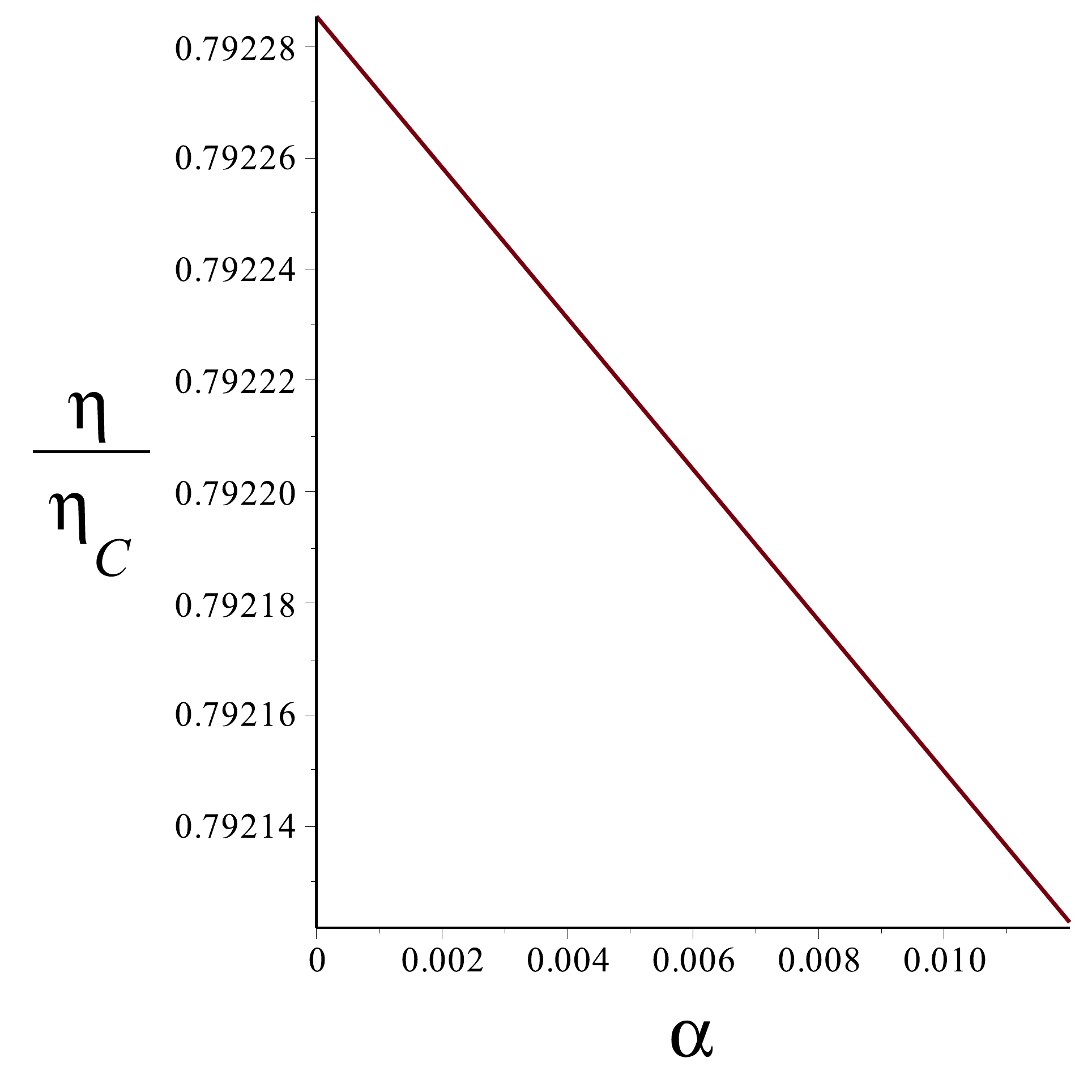} }\hspace{1.5cm}
\subfigure[]{\includegraphics[width=2.5in]{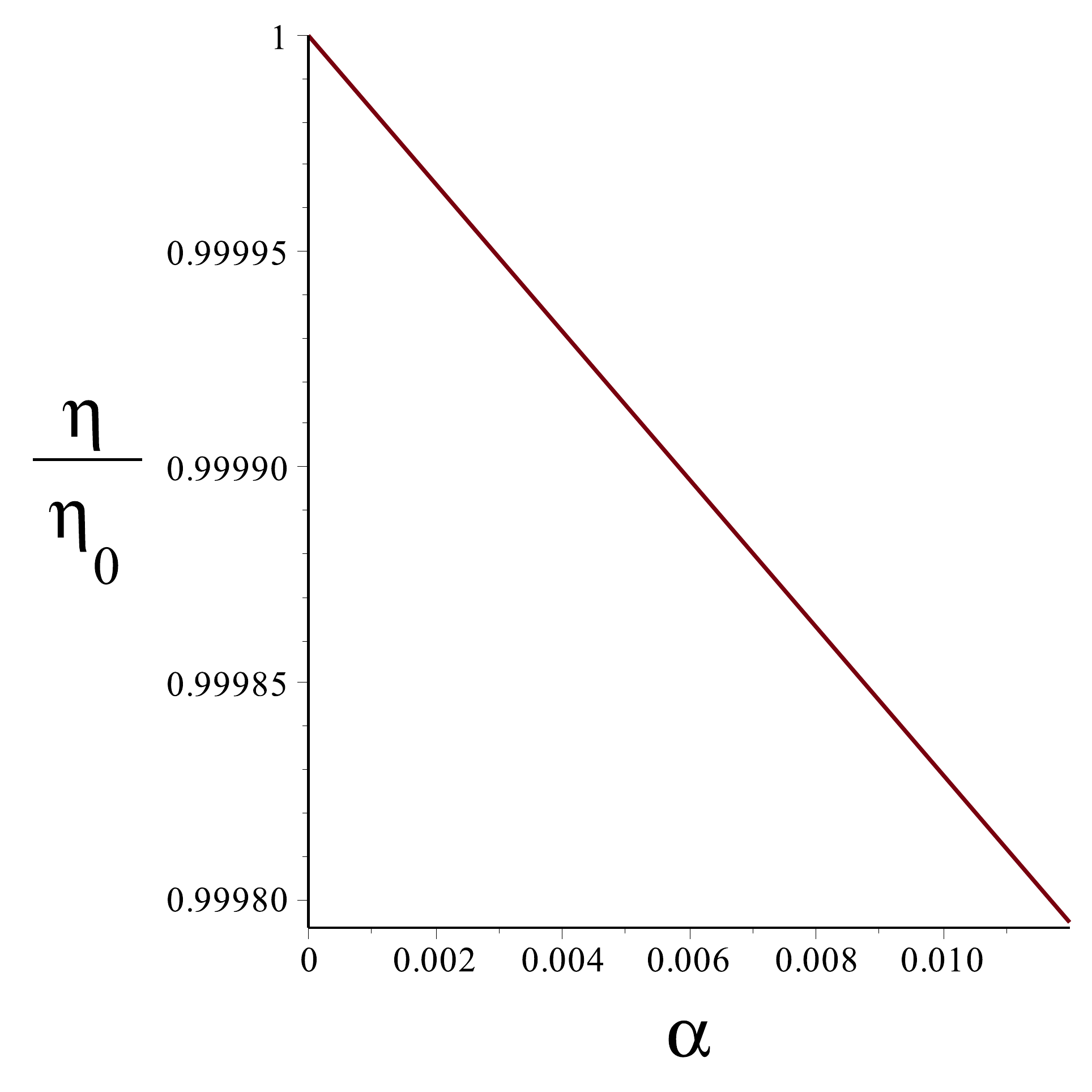}} 
   \caption{\footnotesize  (a) The ratio $\eta/\eta_C$ {\it vs} $\alpha$, and  (b) the ratio $\eta/\eta_0$  {\it vs} $\alpha$ in scheme 2 over the physical range given by equation~(\ref{eq:constrain-alpha}), implemented as described in the text.  This is for $D=5.$ (See the caption of  figure~\ref{fig:efficiency-compare-3} for parameter values.)}   \label{fig:efficiency-compare-4}
}
\end{figure}

\begin{figure}[h]
{\centering
\subfigure[]{\includegraphics[width=2.5in]{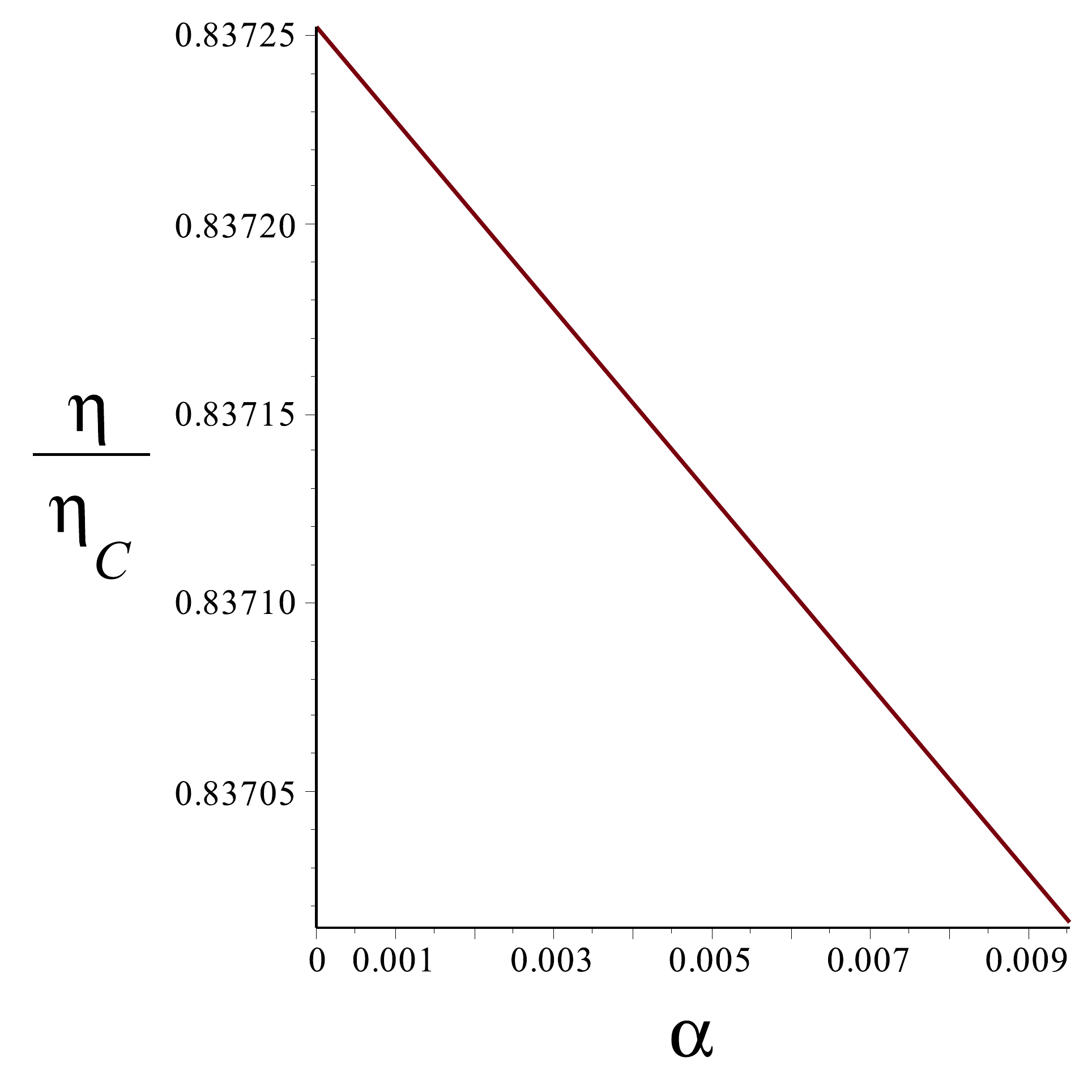} }\hspace{1.5cm}
\subfigure[]{\includegraphics[width=2.5in]{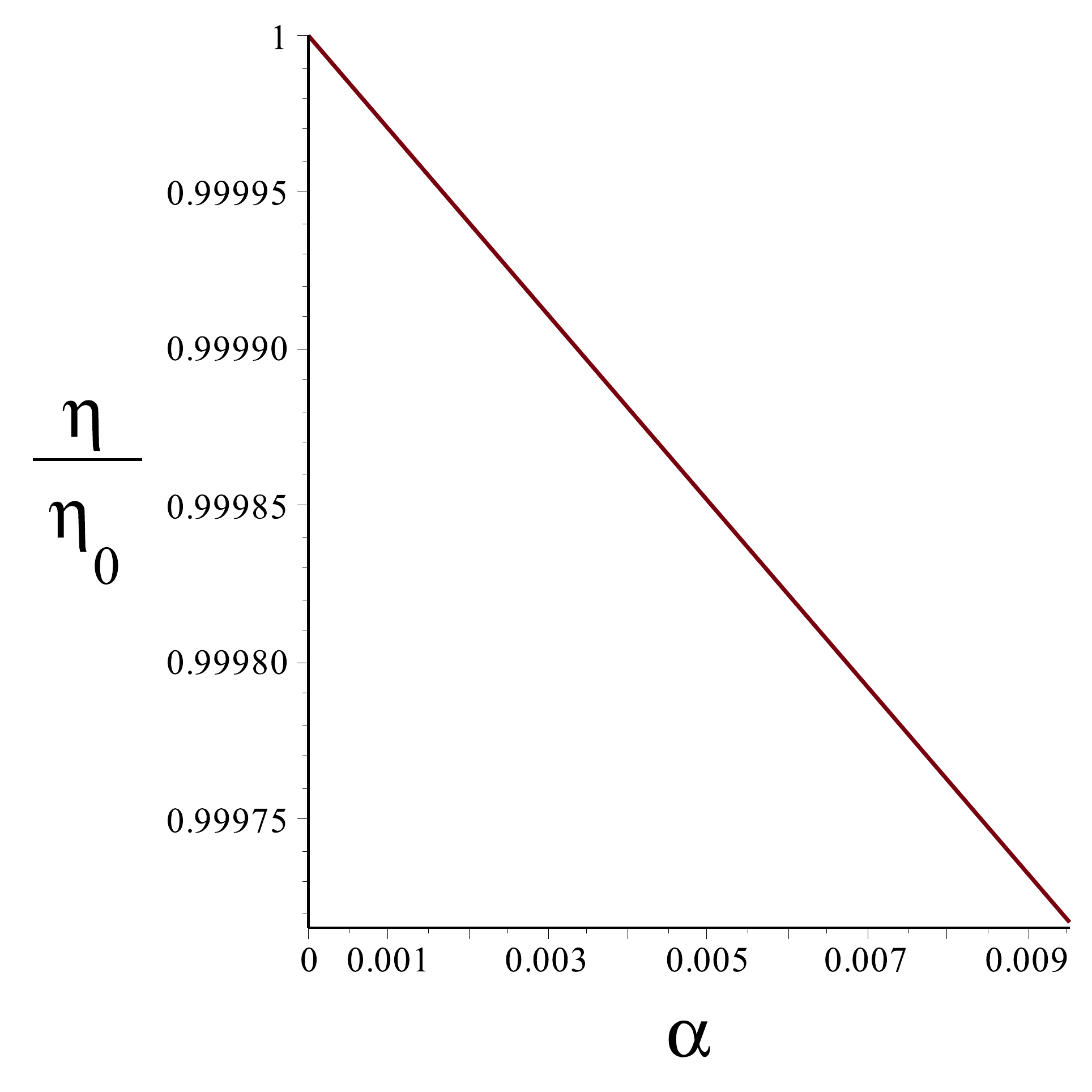}} 
   \caption{\footnotesize  (a) The ratio $\eta/\eta_C$ {\it vs} $\alpha$, and  (b) the ratio $\eta/\eta_0$  {\it vs} $\alpha$ in scheme 2 over the physical range given by equation~(\ref{eq:constrain-alpha}), implemented as described in the text.  This is for $D=6.$ (Again we used the same defining values of the state variables that were used for $D=5$; they are listed in the  caption of  figure~\ref{fig:efficiency-compare-3}.)}   \label{fig:efficiency-compare-5}
}
\end{figure}

Again, we define $\eta^{\phantom{9}}_0$ as the efficiency at $\alpha=0$. Examples of the ratios $\eta/\eta^{\phantom{C}}_{\rm C}$ and $\eta/\eta^{\phantom{C}}_0$ are plotted for $D=5$ in figures~\ref{fig:efficiency-compare-4}(a) and~\ref{fig:efficiency-compare-4}(b), both {\it decreasing}  with~$\alpha$, for a range of sample parameters in contrast with scheme~1. For $D=6$ we plot the analogous quantities in figures~\ref{fig:efficiency-compare-5}(a) and~\ref{fig:efficiency-compare-5}(b). The decreases here also are apparently robust. 

As in scheme~1, this behaviour can be understood from the large $T$ expressions given in equations~(\ref{eq:5dexpansions-alpha}) and~(\ref{eq:6dexpansions-alpha}) by truncating to just the positive powers of $T$. In either $D=5$ or $D=6$ the story is the same. In the case of $D=5$  the efficiency is the product form:
\begin{equation}
\eta=\left(1-\frac{p_4}{p_1}\right)(V_2-V_1)\left\{\frac{81}{512}\pi^2\frac{(T_2^4 -T_1^4)}{p_1^2}+\frac{9}{128}(32\pi\alpha p_1-3)\frac{(T_2^2-T_1^2)}{p_1}
\right\}^{-1}\ ,
\label{eq:productform}
\end{equation}
and there is a similar expression for $D=6$ with the term in braces replaced by the expression
\begin{equation}
\frac{8}{15}\pi^2\frac{(T_2^5 -T_1^5)}{p_1^3}+\frac{4}{3}(4\pi\alpha p_1-1)\frac{(T_2^3-T_1^3)}{p_1^2}\ .
\label{eq:productformalt}
\end{equation}

In both cases,  the factor of the volume difference is fixed in this scheme and so will not contribute any change with $\alpha$.
In this scheme, $T_1$, $p_1$ and $p_4$ must be determined by use of the equation of state, and  some  algebra yields for the terms in braces:
\begin{eqnarray}
&&\left\{\cdot\right\}= \frac{9}{32}T_2\left[\pi 2^\frac12 T_2\frac{(V_2-V_1)}{V_2^\frac12} - \pi^\frac12 2^\frac14\frac{ (V_2^\frac12-V_1^\frac12)}{V_2^\frac14} \right]\nonumber\\
&&\hskip 3.8cm+\frac{9\pi}{32}T_2\left[4\pi T_2\frac{(V_2-V_1)}{V_2} - \pi^\frac12 2^\frac34\frac{ (V_2^\frac12-V_1^\frac12)}{V_2^\frac34} \right] \alpha+O(\alpha^2)\ ,
\end{eqnarray}
for $D=5$, and for $D=6$:
\begin{eqnarray}
&&\left\{\cdot\right\}= \frac{2^\frac15}{15^\frac35}T_2\left[2\pi^\frac45 15^\frac15 T_2\frac{(V_2-V_1)}{V_2^\frac25} - 5\pi^\frac15 2^\frac35\frac{ (V_2^\frac35-V_1^\frac35)}{V_2^\frac15} \right]\nonumber\\
&&\hskip 2.0cm+\frac{8\pi}{15}T_2\left[ 2^\frac75\pi^\frac35 15^\frac15 T_2\frac{(V_2-V_1)}{V_2^\frac45} + \frac{ (10V_2^\frac35-15V_1^\frac15V_2^\frac25+5V_1^\frac35)}{V_2^\frac35} \right] \alpha+O(\alpha^2)\ .
\end{eqnarray}

Since we're working at large $T$, we can ignore the second term in each of the square braces that appear, as the $T_2^2$ terms will dominate. So again, in both cases, we see that the  $\alpha$ coefficient is again positive, to leading order. So in either $D=5$ or $D=6$, the $\{\cdot\}$ term will increase (corresponding to an increase in input heat needed for the cycle), so $\eta$ potentially decreases (we are dividing by  the factor in braces).  This leaves the first factor of the product form~(\ref{eq:productform}) (with~(\ref{eq:productformalt})) of the efficiency.  Use of the equation of state gives at large $T$:
\begin{equation}
\frac{p_4}{p_1}=\frac{T_4}{T_2}\left(\frac{V_2}{V_1}\right)^\frac14\left(1+\sqrt{2}\pi\left(\frac{V_2^\frac12-V_1^\frac12}{V_2^\frac12 V_1^\frac12}\right)\alpha+O(\alpha^2)\right)\ ,
\end{equation}
for $D=5$ and
\begin{equation}
\frac{p_4}{p_1}=\frac{T_4}{T_2}\left(\frac{V_2}{V_1}\right)^\frac15\left(1+\frac{4\pi^\frac45 2^\frac15}{(15)^\frac25}\left(\frac{V_2^\frac25-V_1^\frac25}{V_2^\frac25 V_1^\frac25}\right)\alpha+O(\alpha^2)\right)\ ,
\end{equation}
for $D=6$. This means that the first factor  in brackets of the product form decreases with $\alpha$ too.  We see therefore that in both $D=5$ and $D=6$, the ratio $\eta/\eta_0$ must decrease (at least in this large~$T$ limit). Since in this scheme, the Carnot efficiency is also fixed at the outset, we see that $\eta/\eta^{\phantom{C}}_{\rm C}$ must also generically decrease (in this large~$T$ limit), again confirming what was uncovered in the numerical exploration.

\section{Closing Remarks}
After careful clarification and extension of the results of ref.\cite{Johnson:2014yja} for holographic heat engines in Einstein--Maxwell gravity (with negative $\Lambda$) we added a Gauss--Bonnet sector and computed analogous results for similarly defined heat engines. It was particularly interesting to see what happens to the efficiency $\eta$ (at least in the large $T$ limit) in the presence of the Gauss--Bonnet sector, for the particular choice of engine cycle defined in figure~\ref{fig:cyclesb}. Such higher curvature corrections to the gravity theory affect both the black holes' heat capacity and their ability to do mechanical work (since their size depends upon the Gauss Bonnet coefficient $\alpha_{\rm GB}$) and hence it is not {\it a priori} obvious as to how the efficiency would be affected.  Moreover, we learned that how $\eta$ is affected also depends upon what parameters of the engine cycle are held fixed as $\alpha_{\rm GB}$, the coefficient of the Gauss--Bonnet terms, is changed. (Recall that the parameter $\alpha$, used in much of the presentation,  is  proportional to $\alpha_{\rm GB}$; see just below equation~(\ref{eq:why}).)

We compared the efficiency as  a function of $\alpha_{\rm GB}$ to two natural quantities: (1)  the efficiency at $\alpha_{\rm GB}=0$, denoted $\eta^{\phantom{0}}_0$, and (2) the maximum (Carnot) efficiency, $\eta^{\phantom{C}}_{\rm C}$. Furthermore, we examined two schemes for which parameters of the engine to hold fixed as we change $\alpha_{\rm GB}$, and the effects on the efficiency differed in each scheme. This is because the equation of state is $\alpha_{\rm GB}$ dependent, and so the parameters of the engine {\it not} held fixed depend upon $\alpha_{\rm GB}$. This  has a scheme dependent effect on the efficiency.  The question therefore arises as to which is the better overall benchmark for measuring the effects of the Gauss--Bonnet sector on $\eta$. If there is anything close to a scheme--independent answer to this question (and it is not clear that there is),  perhaps the Carnot standard, asking how $\eta/\eta^{\phantom{C}}_{\rm C}$ has changed, is more robust since there is an unambiguous and universal meaning to what that Carnot engine is, and its efficiency takes as input only the highest and lowest temperatures ($T_H$ and $T_{\rm C}$ respectively) in the engine under discussion. This becomes even stronger in scheme  2 which holds $T_H$ and $T_{\rm C}$ fixed as $\alpha_{\rm GB}$ varies.

The physics of Einstein gravity with negative cosmological constant  is believed to be holographically dual to large $N$ (for suitably defined $N$) field theories in one dimension fewer, and higher curvature corrections such as Gauss--Bonnet  terms correspond to particular classes of $1/N$ corrections to the duality (for a review, see ref.\cite{Aharony:1999t}). While it is not yet clear what the extended thermodynamics (and in particular, our holographic  heat engines)  might teach us about the dual field theories, some suggestions about how to proceed have been made in the literature:  Since varying $p$ ends up changing the $N$ of the field theory ({\it e.g.} the $N$ in an $SU(N)$ Yang--Mills in the case of $D=5$), a heat engine really defines a cycle on the space of the dual field theories\cite{Johnson:2014yja}, since~$N$ is not fixed\footnote{Changing $p$ (and hence $l$) also changes the volume the dual field theory resides on, as pointed out in ref.\cite{Karch:2015rpa}.}. As suggested in ref.\cite{Johnson:2016pfa}, it is possible that the efficiency of an holographic heat engine, involving heat flows and the performance of mechanical work as it does, might  characterise an important physical property of the field theory (or class of field theories) to which the engine is dual, such as a near--equilibrium expansion of an appropriate response function. Of course, further work (beyond the scope of this paper) is needed to sharpen this suggestion.

Whatever the meaning of the efficiency for the dual large~$N$ field theories, our  results  tell us about the changes to that quantity once this specific class of  $1/N$ corrections are added: At least in a high $T$ expansion, when compared to $\eta^{\phantom{C}}_{\rm C}$ or $\eta_0=\eta(\alpha=0)$ there is a clear {\it decrease} in $\eta$ with increasing $\alpha$ in scheme~2, at least for $D=5$ and $D=6$.  The ratio $\eta/\eta_0$ also has a generic {\it increase} with $\alpha$ in scheme~1, for both $D=5$ and $D=6$. (The sign of the slope of $\eta/\eta^{\phantom{C}}_{\rm C}$ is not robust  in scheme~1, since the  lowest temperature of the engine is itself parameter dependent.)

It would be interesting to establish what happens in higher dimensions, and to examine the physics away from the high $T$ regime. The recent results of ref.\cite{Johnson:2016pfa}  may be useful in that regard. It would also be interesting to study the effects on the heat engines that can be defined for  other classes of extension or deformation of Einstein--Maxwell gravity, whether motivated by dual field theory considerations or not. We hope to report new results from this avenue of investigation in the near future.

\section*{Acknowledgements}
 CVJ would like to thank the  US Department of Energy for support under grant DE-FG03-84ER-40168,  and Amelia for her support and patience.  Further research included in a revision of this manuscript was done at the Aspen Center for Physics, supported in part by the National Science Foundation grant PHY-1066293.


\providecommand{\href}[2]{#2}\begingroup\raggedright\endgroup

\end{document}